\documentclass[sn-mathphys,Numbered]{sn-jnl}% Math and Physical Sciences Reference Style
\usepackage{graphicx}%
\usepackage{multirow}%
\usepackage{amsmath,amssymb,amsfonts}%
\usepackage{amsthm}%
\usepackage{mathrsfs}%
\usepackage[title]{appendix}%
\usepackage{xcolor}%
\usepackage{textcomp}%
\usepackage{manyfoot}%
\usepackage{booktabs}%
\usepackage{algorithm}%
\usepackage{algorithmicx}%
\usepackage{algpseudocode}%
\usepackage{listings}%
\usepackage{adjustbox}
\usepackage[normalem]{ulem}
\theoremstyle{thmstyleone}%
\newtheorem{theorem}{Theorem}%  meant for continuous numbers
\newtheorem{proposition}[theorem]{Proposition}% 

\theoremstyle{thmstyletwo}%
\newtheorem{remark}{Remark}%

\theoremstyle{thmstylethree}%

\begin{document}

\title[A mathematical model for fibrous dysplasia]{A mathematical model for fibrous dysplasia: The role of the flow of mutant cells}


\author*[1]{\fnm{Mariia} \sur{Soloviova}}\email{mariia.soloviova@uclm.es}

\author[1]{\fnm{Juan Carlos} \sur{Beltran Vargas}}

\author[2]{\fnm{Luis} \sur{Fernandez de Castro}}

\author[1]{\fnm{Juan} \sur{Belmonte-Beitia}}

\author[1]{\fnm{V\'{\i}ctor M.} \sur{P\'erez-Garc\'{\i}a}} 

\author[3]{\fnm{Magdalena} \sur{Caballero}} 

\affil*[1]{\orgdiv{Department of Mathematics, Mathematical Oncology Laboratory (MOLAB)}, \orgname{Universidad de Castilla-La Mancha}, \orgaddress{\street{Avda. Camilo Jos\'e Cela 3}, \city{Ciudad Real}, \postcode{13071}, \country{Spain}}}

\affil[2]{\orgdiv{Skeletal Biology Section, National Institute of Dental and Craniofacial Research, National Institutes of Health, Department of Health and Human Services}, \orgname{National Institutes of Health}, \orgaddress{\street{} \city{Bethesda MD}, \postcode{} \country{USA}}}

\affil[3]{\orgdiv{Department of Mathematics}, \orgname{Universidad de C\'ordoba}, \orgaddress{\street{Campus de Rabanales} \city{C\'ordoba}, \postcode{14071},\country{Spain}}} 

\abstract{Fibrous dysplasia (FD) is a mosaic non-inheritable genetic disorder of the skeleton in which normal bone is replaced by structurally unsound fibro-osseous tissue. There is no curative treatment for FD, partly because its pathophysiology is not yet fully known. We present a simple mathematical model of the disease  incorporating its basic known biology, to gain insight on the dynamics of the involved bone-cell populations, and shed light on its pathophysiology. We develop an analytical study of the model and study its basic properties. The existence and stability of steady states are studied, an analysis of sensitivity on the model parameters is done, and different numerical simulations provide findings in agreement with the analytical results. We discuss the model dynamics match with known facts on the disease, and how some open questions could be addressed using the model.}



\maketitle

\section{Introduction}\label{sec1}

Fibrous dysplasia (FD) is a mosaic non-inheritable genetic disorder of the skeleton in which normal bone is replaced by structurally unsound fibro-osseous tissue. FD is occasioned by post-zygotic activating mutations in the gene \emph{GNAS} which encodes the $\alpha$-subunit of the Gs stimulatory protein (G$\alpha$s) \cite{Zhao2018}, resulting in an inappropriate overproduction of intracellular cyclic adenosine monophosphate (cAMP) \cite{Hartley2019}. During embryonic development, mutant progeny cells migrate, resulting in a mosaic distribution and leading to a broad spectrum of disease burden.
In bone, G$\alpha$s activation causes altered differentiation of skeletal stem cells (SSC). Lesional osteoprogenitors proliferate excessively and adopt a fibroblastic phenotype, forming a fibro-osseous tissue with variable presence of abnormal curvilinear trabecules of woven bone, leading to discrete skeletal lesions prone to fracture, deformity, and pain \cite{Boyce2020}. Lesions become visible in early childhood, and expand to reach final burden in early adulthood, when lesions stabilize and stop expanding \cite{Hart2007}. FD may appear alone or in association with extraskeletal manifestations, including hyperpigmented macules and hyperfunctioning endocrinopathies. The combination of two or more of the previous features is named McCune-Albright syndrome (MAS) \cite{Collins2012}. 

There is no cure for FD, and there are no known effective medical therapies or treatments. Surgeries to preserve function 
and reduce pain are performed, combined with physical therapy, orthoses, avoidance of prolonged immobilization, and management of 
underlying endocrinopathies \cite{Paul2014}. Antiresorptive therapies, including denosumab and bisphosphonates, are being studied to treat FD, \cite{Chapurlat2021,Collins2020,deCastro2023}. These therapies inhibit osteoclast function modifying the imbalance of osteoblasts and osteoclasts. However, there are no clear guidelines for the dosage and scheduling that each patient should receive. 

The complete understanding of the progression rate of FD lesions and the biochemical factors that instigate their formation remains \cite{Szymczuk2022}. This lack of clarity presents challenges in investigating and implementing preventive therapies. The pathophysiology of FD is not fully grasped, and recent works \cite{Whitlock2023-2} also highlight undisclosed aspects of bone remodeling physiology. For instance, the intricate coordination and communication between osteoclasts and osteoblasts lack a comprehensive mechanistic understanding, presenting a substantial barrier to comprehending the biology governing bone remodeling and developing effective treatments for this pivotal process.

Bone biology and different aspects of its remodelling have been previously considered by mathematical biologists \cite{PivonkaKomarova2010}. Mathematical models, typically based on ordinary differential equation (ODEs), of healthy bone, injured bone, and bone disorders different from FD have been constructed previously \cite{Komarova2003,Lemaire2004,Komarova2005,Pivonka2008,Ayati2010,Buenzli2012,Graham2013,Lo2021}. However, to our knowledge no previous mathematical models of FD have been constructed.

In this study, we construct and analyze a mathematical model to depict the cellular dynamics within FD tissue. The model captures the temporal evolution of various interacting bone-cell populations, including altered and normal osteoprogenitors, alongside osteoblasts, osteocytes, and osteoclasts, focusing on young adults during the stabilization phase when lesions cease to expand. These cell populations are averaged over a sufficiently large bone volume to ensure consistent outcomes. Our objective is to identify the biological processes that govern the composition of lesions in both general bone remodeling and specifically in FD. Additionally, we have developed a model for healthy bone tissue to serve as a tool for determining the necessary parameters for our FD model, incorporating the biological alterations seen in FD.

This article is structured as follows: In Section 2, we establish a mathematical model for both FD and healthy bone. Section 3 focuses on the estimation of model parameters. Section 4 presents the principal mathematical findings, including existence, boundedness, steady states, stability, and various numerical simulations for both models, providing valuable biological insights into their dynamics. Additionally, a sensitivity analysis is conducted to discern the crucial model parameters. Finally, in Section 5, we recapitulate our results and engage in a discussion concerning the alignment between the mathematical outcomes and the underlying biology of the disease. 

\section{The model}
\label{model}

\subsection{Biological foundations}

Bone remodeling is a process occurring at specific spatial and temporal locations in the skeleton, orchestrated by the coordinated actions of osteoclasts and osteoblasts organized into basic multicellular units (BMUs) \cite{Frost1964, Bolamperti2022}. In adults, approximately 5\%-10\% of the skeleton undergoes replacement annually, resulting in a complete turnover of the entire skeleton within a 10-year period \cite{Parfitt1980, Sims2014}. The forthcoming mathematical models will account for the dynamic evolution over time of various interacting bone-cell populations, averaged over a sufficiently large bone volume. Consequently, the values of each population will represent spatial averages across a finite volume of bone containing numerous BMUs, distinguishing our model from those that consider a single BMU \cite{Arias2018}.

The mechanisms and timing of the remodeling process vary between cortical and trabecular bone \cite{Eriksen2010, Sims2014}. Fibrous dysplasia lesions manifest in trabecular bone, extending into the bone marrow space and causing cortical thinning \cite{Boyce2020}. Consequently, our model will specifically address the characterization of trabecular bone and the modifications induced by the disease.

In healthy young adults, remodeling events involve the replacement of resorbed bone with an equivalent amount of new bone. During childhood and adolescence, a continuous gain of trabecular bone mass occurs through a combination of remodeling and modeling events, reaching a peak in early adulthood. Subsequently, both men and women experience trabecular bone loss starting in the third decade of life \cite{Khosla2012}. The behavior of fibrous dysplasia lesions is strongly age-dependent, exhibiting heightened activity in children and teenagers, stabilization in young adults, and, in older patients, partial or complete normalization of histology \cite{Kuznetsov2008, Florenzano2019}. We have opted to model the remodeling process occurring in young adults, a phase where significant therapeutic interventions may be necessary to manage the disease or alleviate its impact.

Additionally, the regulation of bone remodeling processes involves both systemic factors, such as hormones, and local components \cite{Hadjidakis2007}. In our simplified depiction, our focus will be on the key processes altered in fibrous dysplasia (FD), and thus, we will exclusively address the local regulatory components.

Bone remodeling is a intricate process governed by the interplay of various chemical signals. However, in this paper, we will not delve into the quantitative intricacies of these signals. Instead, our attention will be directed solely towards the bone-cell populations emitting and receiving these signals. Consequently, we will forego quantifying the signals that facilitate interactions between the populations, opting to represent them as a direct impact of specific populations on others.

Remodeling events in healthy adult bone are typically initiated by the apoptosis of osteocytes in response to either mechanical or chemical stimuli \cite{Komori2016}. Those events are followed by a decrease in sclerostin levels and other factors that promote the differentiation of precursors into osteoblasts, \cite{Baron2013}. The same events are also responsible for the increase of the levels of nuclear factor kappa-B ligand (RANKL) in the bone marrow, promoting the recruitment of osteoclasts precursors and its posterior differentiation, \cite{Boyce2007,Nakashima2011}. Cells from the osteoblastic lineage also contribute to osteoclastogenesis through the production of macrophage colony-stimulating factor \cite{Collin-Osdoby2001}, RANKL \cite{Streicher2017}, and other coupling factors \cite{Maeda2012,Kobayashi2015}. Mature osteoclats resorb bone, and are followed by reversal cells (osteoblast progenitors that digest fibrillar collagen remnants) and osteoblasts that deposit new bone, \cite{Bolamperti2022}. When the remodeling process finishes, osteoblasts destiny is any of the following three: undergoing apoptosis (60-80\%), becoming lining cells (cells that cover quiescent bone surfaces that are not undergoing remodeling, which can also be osteblasts precursors), or being trapped in the bone matrix, mineralizing it and differentiating into osteocytes (5-20\%) \cite{Manolagas2010, Bonewald2011, Zhao2000}.

FD lesions are characterized by the presence of osteoprogenitors carrying a mutation in the \emph{GNAS} gene, to be denoted hereafter as mutant osteoprogenitors. These abnormal osteoprogenitors overproduce cAMP, \cite{Riminucci1997}, resulting in a cAMP related reversible cell phenotype of surrounding wild-type (WT) osteoprogenitors, \cite{Xiao2019}, to be referred hereafter as WT phenocopying osteoprogenitors. Mutant osteoprogenitors, as well as WT phenocopying ones, proliferate at a higher rate and differentiate abnormally, producing aberrant abnormal woven bone and fibrous matrix, and releasing factors to induce osteoclastogenesis and bone resorption \cite{Marie1997, Riminucci2003, Zhao2018, Hartley2019}. The final balance between resorption and formation of bone is therefore altered.

\begin{figure}[H]
	\centering
	
	\includegraphics[width=0.75\textwidth]{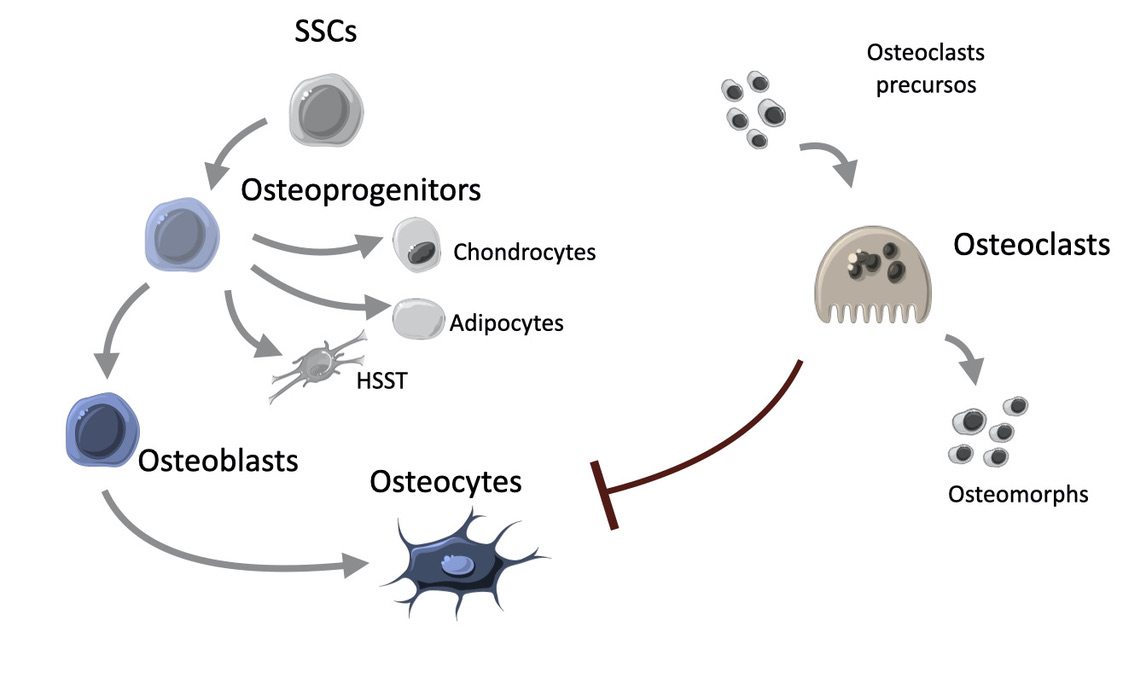} \\
	\textbf{(a)}   \\[6pt]
	\includegraphics[width=0.95\textwidth]{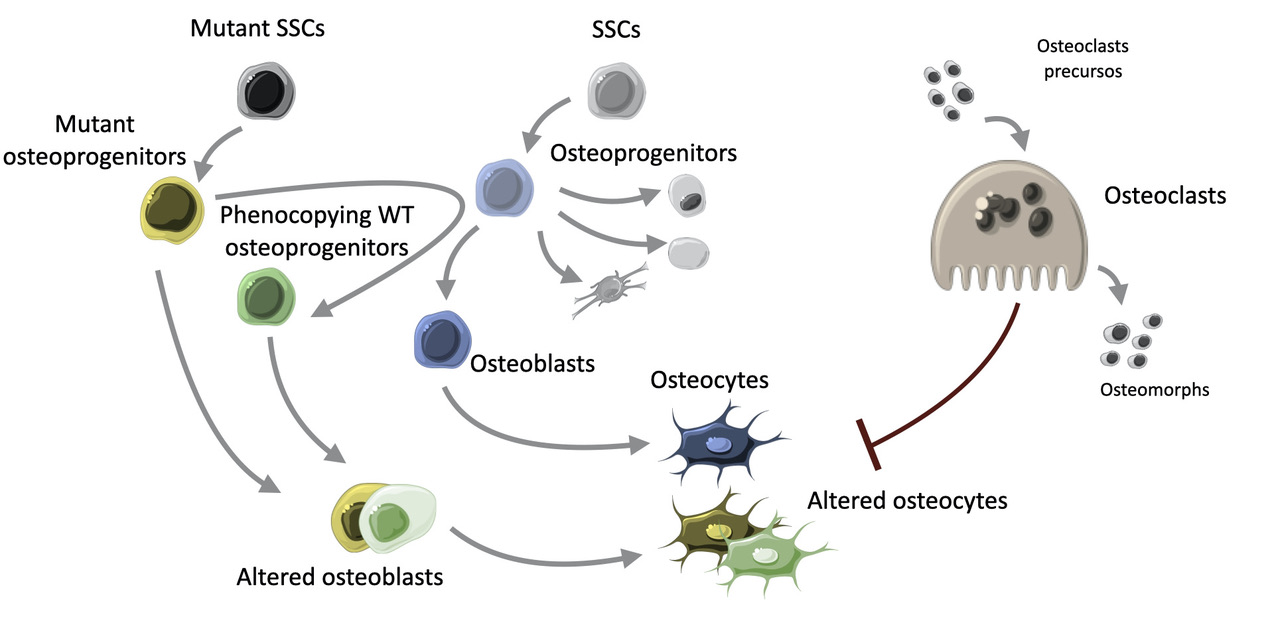} \\
	\textbf{(b)} \\[6pt]

\caption{\textbf{Bone-cell populations relevant in bone remodeling, for both healthy bone and FD tissue.} In (a), bone-cell populations involved in bone remodeling are shown. Bone remodeling consists of three coupled processes: resorption, deposition, and mineralization. The first of these processes is carried out by osteoclasts, whereas the last two are performed by cells of the osteoblastic lineage. 	
	Osteoclasts are large multinucleated bone-resorbing cells formed by the fusion of monocyte/macrophage-derived precursors that reside in the bone marrow.  Osteoclastogenesis is driven by signals coming from osteocytes and osteoprogenitors. In healthy bone, fusion can only occur at the bone surface to be resorbed. Mature activated osteoclasts have a finite lifetime and after resorption, osteoclasts either undergo apoptosis (a rare event with a high-energy cost due to the removal of the apoptotic debris) or disassembly into mononucleated cells unable to resorb bone, osteomorphs, that remain in the adjacent bone marrow, \cite{Bolamperti2022,McDonald2021}.
The osteoblastic lineage derives from the differentiation of skeletal stem cells (SSCs) in the bone marrow. SSCs are self-renewing multipotent progenitors that can give rise to osteoblasts (bone-depositing cells), chondrocytes, hematopoietic-supportive stromal cells, and marrow adipocytes, \cite{Bianco2015}. 
When the remodeling process finishes, osteoblasts destiny is any of the following three: undergoing apoptosis (60-80\%), becoming lining cells (cells that cover quiescent bone surfaces that are not undergoing remodeling, which can also be osteblasts precursors), or being trapped in the bone matrix (mainly consisting of collagen), mineralizing it and differentiating into osteocytes, \cite{Manolagas2010, Bonewald2011, Zhao2000}. Osteocytes are terminally differentiated cells that represent more than 90\% of
	bone cells in the adult skeleton, and they are considered to be the orchestrators of the bone remodeling, \cite{Bolamperti2022,Bonewald2011}.	In (b), bone-cell populations appearing on FD tissue are added.  Mutant SSCs differentiate into mutant osteoprogenitors which overproduce cAMP. Surrounding WT osteoprogenitors adopt a cAMP related reversible cell phenotype, being called WT phenocopying osteoprogenitors. Both mutant and WT phenocopying osteoprogenitors overproliferate and differentiate abnormally, giving rise to altered osteoblasts and osteocytes, and producing aberrant abnormal woven bone and fibrous matrix \cite{Riminucci1997,Marie1997,Xiao2019,Zhao2018,Hartley2019}.	They also release  factors to induce osteoclastogenesis and bone resorption, causing the formation of ectopic, numerous, large (and so, more active) osteoclasts, \cite{Riminucci2003,Whitlock2023}. }
\label{fig:Cells}
\end{figure}

A schematic summary of these processes is illustrated in Figure \ref{fig:Cells}.

\newpage
\thispagestyle{empty}
\subsection{FD mathematical model}

\begin{figure}[H]
	\centering
	\includegraphics[width=0.95\textwidth]{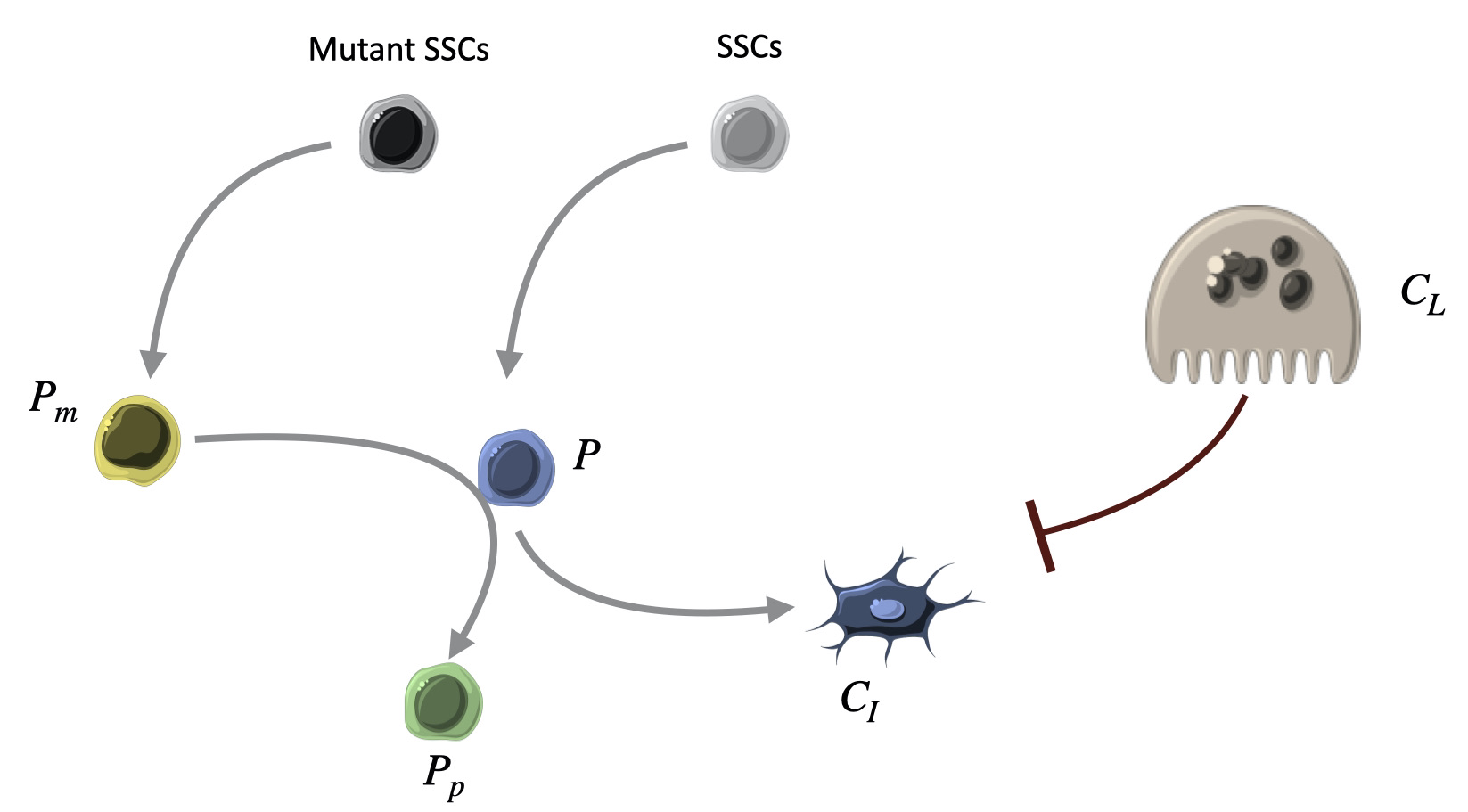} \\
	\textbf{(b)} \\[6pt]
	
	\caption{\textbf{Schematic representation of the bone-cell populations and interactions between them considered in the FD mathematical model given by Eqs. \eqref{FD_simple_model}.} Bone remodeling consists of three coupled processes: resorption, deposition, and mineralization. For each of those processes, we have chosen a bone-cell population that accomplishes the process: osteoclasts ($C_L$), WT osteoprogenitors ($P$), and mature osteocytes ($C_I$). For the sake of simplification, both osteoprogenitors and osteoblasts (as well as linning and reversal cells) are clustered into a unique population, which by abuse of notation will also be referred to as osteoprogenitors. Since we are considering FD tissue, we have added two more populations, mutant and WT phenocopying osteoprogenitors ($P_m$ and $P_p$, respectively). They give rise to fibrous tissue rich in fibroblast-like cells that express markers of early stages of osteogenic maturation. For this reason, in our FD model, the mutant osteoprogenitor population also comprehends mutant osteoprogenitors progeny, and the same applies for the WT phenocopying osteoprogenitor population. }
	\label{fig:Cells2}
\end{figure}

Let $P(t)$,
$P_m(t)$, $P_p(t)$, $C_I(t)$, and $C_L(t)$ denote non-negative time-varying functions
representing the number of osteoprogenitors cells (including osteoblasts, lining cells, and reversal cells), mutant osteoprogenitors (includying its progeny), WT phenocopying osteoprogenitors (including its progeny), mature osteocytes, and osteoclasts, respectively, normalized such that the cellular level at which osteoprogenitors stop proliferating is the unity. SSCs are assumed to be constant, and so are not
modeled explicitly. See Figure \ref{fig:Cells2}.

Our initial autonomous system of differential equations of compartmental type describing FD is as follows:
\begin{subequations}
\label{FD_simple_model}
\begin{eqnarray}
    \cfrac{dP}{dt}&= & \rho P (1 -  M) + \tau_s (1-\mu M) - \alpha P - \delta_m P P_m,  \label{FD_simple_model1}\\
    \cfrac{dP_m}{dt}&= & \rho_m P_m (1 -  M) + \tau_m (1-\mu M), \label{FD_simple_model2}\\
    \cfrac{dP_p}{dt}&= & \rho_m P_p (1 -  M) + \delta_m P P_m, \label{FD_simple_model3}\\
    \cfrac{dC_I}{dt}&= & \alpha P - \delta C_I C_L,\label{FD_simple_model4}\\
    \cfrac{dC_L}{dt}&= &  \tau_c M - \gamma_c C_L,\label{FD_simple_model5}\\
    M &=  & P+P_m+P_p+C_I. \label{FD_simple_model6}%\\
\end{eqnarray}
\end{subequations}

Equation \eqref{FD_simple_model1} describes the time dynamics of the osteoprogenitor population, that is the population mediating the formation of new bone. The first term in Eq. \eqref{FD_simple_model1} accounts for the factual proliferation rate, taking a logistic form accounting for limitations of growth due to competition for space in the trabeculas. For young adults, the volume of trabeculas is conserved by bone remodeling, \cite{Khosla2012}. Therefore, the more bone is resorbed, the more new bone needs to be deposited, \cite{Bolamperti2022}. And so, osteoclasts are indirectly included in the saturation term $M$ via the number of osteocytes. 
 It is known that at all differentiation stages, osteoprogenitors undergo apoptosis, \cite{Komori2016}. Factual proliferation comprehends this phenomenon. Indeed, when this first term of the equation becomes negative, it means that apoptosis is surpassing proliferation.
 
The second term in Eq. \eqref{FD_simple_model1} describes the flow of osteoprogenitors differentiating from SSCs. The flow is modulated by the expression $(1-\mu M)$. Its biological explanation relies on the fact that the receptor
 activator of nuclear factor kappa-B (RANK) is expressed in SSCs, and when RANKL binds to it, the RANKL forward signaling is activated, inhibiting osteogenic differentiation, \cite{Chen2018}. Whereas the first term of this equation can be negative, that is not the case of the second term, since it has no biological meaning. Therefore, the parameter $\mu$ needs to be added to have a biologically reasonable behavior. The third term describes the natural process of differentiation of osteoprogenitor cells into mature osteocytes, a process assumed to have a characteristic time  $1/\alpha$ \cite{Bonewald2011}.  Finally, the last term in Eq. \eqref{FD_simple_model1} corresponds to the change of WT osteoprogenitor cells into WT phenocopying osteoprogenitors.  The parameter $\delta_m$ measures the probability (per unit of time) of an encounter of WT osteoprogenitors and mutant osteoprogenitors. Mutant osteoprogenitors overproduce cAMP, \cite{Riminucci1997}. Expelled cAMP is absorbed by surrounding WT osteoprogenitors, resulting in a cAMP related reversible cell phenotype, \cite{Xiao2019}. Those cells constitute the WT phenocopying osteoprogenitor population. The increase of $P_p$ due to this process is proportional to the amount of cells from $P$ that surround cells from $P_m$. As a simplification, we consider that all the cells from $P$ are close enough to cells from $P_m$.

Equations \eqref{FD_simple_model2} and \eqref{FD_simple_model3} describe the dynamics of mutant and WT phenocopying osteoprogenitor cells. The first term takes into account the proliferation rate of this kind of cells, assumed to be similar for both cellular subpopulations. Since both populations proliferate faster than normal osteoprogenitors, it should be expected that $\rho_m > \rho$, \cite{Riminucci1997, Hartley2019}. These cells have also been found to be highly apoptotic, \cite{Kuznetsov2008}.   

The second term in Eq. \eqref{FD_simple_model2} accounts for the flow of osteoprogenitors differentiating from mutant SSCs, \cite{Zhao2018}. Note than there is no term for differentiation of altered osteoprogenitors into mutant osteocytes, neither in Eq.  \eqref{FD_simple_model2} nor in  \eqref{FD_simple_model3}  because, due to their alterations, these cells cannot mature properly. 

To complete the equations for the osteoprogenitor compartments, the last term in Eq.  \eqref{FD_simple_model3} accounts for normal osteoprogenitors acquiring the mutant phenotype and matches the last term in Eq. \eqref{FD_simple_model1} 

As to the equation describing the dynamics of the osteocyte population \eqref{FD_simple_model4}, its first term accounts for the maturation of normal osteocytes, mirroring the third term in Eq.  \eqref{FD_simple_model1}. We also account for the resorption of osteocytes by osteoclasts in the second term in Eq. \eqref{FD_simple_model4}. The parameter $\delta$ measures an effective probability (per unit of time) of an encounter between osteoclasts and osteocytes leading to osteocyte resorption.

Our last equation, \eqref{FD_simple_model5}, delineates the dynamics of osteoclasts. Osteoclastogenesis is instigated by signals emanating from osteocytes and osteoprogenitors \cite{Collin-Osdoby2001,Streicher2019,Yang2019,Xiong2011}, and as such, the structure of the first term captures this effect. Mature activated osteoclasts have a finite lifespan, and following bone resorption, they either undergo apoptosis—a rare event incurring a high-energy cost due to the removal of apoptotic debris—or disassemble into smaller cells unable to resorb bone, termed osteomorphs, which persist in the adjacent bone marrow \cite{McDonald2021}. We have incorporated the decline of osteoclasts through a straightforward elimination term, corresponding to an effective half-life of approximately $1/\gamma_c$. A more detailed exploration of the biological effects beyond this osteoclast clearance term will be provided later.

Finally, in Eqs. (\ref{FD_simple_model1}-\ref{FD_simple_model3}), we have chosen a logistic model to  limit growth in a standard way, although other growth functions could be considered here.

\subsection{Healthy bone model}

To obtain a simplified model of healthy bone behavior we just set $P_m = P_p =0$ in Eqs. \eqref{FD_simple_model} to obtain
\begin{subequations}
\label{HB_simple_model}
\begin{eqnarray}
    \cfrac{dP}{dt}&= & \rho P (1- M) + \tau_s (1-\mu M) - \alpha P,  \label{HB_simple_model1}\\
    \cfrac{dC_I}{dt}&= & \alpha P - \delta C_I C_L,\label{HB_simple_model2}\\
    \cfrac{dC_L}{dt}&= &  \tau_c M - \gamma_c C_L,\label{HB_simple_model3}\\
    M &=  & P+C_I. \label{HB_simple_model4} 
\end{eqnarray}
\end{subequations}

This is quite a simplified model for healthy bone but incorporates the basic compartments of osteoprogenitors, osteocytes and osteoclasts and the basic interactions between them. In contrast to previous modelling approaches \cite{Komarova2003, Komarova2005,  Lemaire2004,  Pivonka2008,  Ayati2010,  Buenzli2012}, our model does not describe a single episode of bone remodeling, but rather what happens on average in a bone segment of sufficiently large size.

The objective of this model for healthy bone is to encompass the critical processes disrupted in fibrous dysplasia (FD). Specifically, we consider the progression of osteoprogenitors differentiating from SSCs (indeed, FD can be viewed as a disorder of postnatal SSCs, leading to the generation of dysfunctional osteoblasts). Additionally, we introduce a logistic term for osteoprogenitors and osteocytes to incorporate growth limitations arising from competition for space. This model will serve as a valuable tool later for estimating certain parameters that may not be readily available in the existing literature.

\section{Parameter estimation}\label{Sec:estimation}

Bone resorption and formation rates are difficult to measure in  vivo, both in humans and in animal models \cite{Lo2021}. Bone is opaque and has a high refractive index, making it difficult to image bone cells in live animals \cite{McDonald2021}. Advances in understanding the bone biology have largely been gained from in vitro studies, but some cell populations behave substantially different in vivo than in vitro \cite{McDonald2021}. Thus, parameter estimation is a common problem in mathematical models of normal bone and bone disorders \cite{PivonkaKomarova2010}.

So we need to estimate the seven parameters appearing in the healthy bone model \eqref{HB_simple_model} and those in the fibrous dysplasia model \eqref{FD_simple_model}. Note that some properties of normal bone are altered in FD, thus some parameters change their value when applied to dysplasic bone.

First of all, the proliferation rate of osteoprogenitors will be estimated using the fact that bone remodeling in trabecular bone takes around $200$ days to complete, \cite{Eriksen2010}. $\rho$ is the parameter on top of the maturation cascade determining the speed of bone reconstruction. From our simulations of Eqs. (\ref{HB_simple_model}), we found that values in the range 0.135-0.165 day$^{-1}$ give the expected bone recovery dynamics. As to the mutant cells, it was found in \cite{Marie1997}, that cell proliferation evaluated by DNA synthesis was two-fold to threefold larger in osteoblastic cells expressing the mutation compared with normal cells from the same patient. Experiments also showed that it was also larger in cells isolated from more severe than less severe fibrotic lesions. Thus, we took $\rho_m\in [2\rho,3\rho]$ to account for a range of aggressiveness in FD lesions.

The osteoprogenitor maturation rate $\alpha$ depends on many factors, but typical residence times in mice is around 5 days \cite{Weng2022}. We have estimated $\alpha = 0.2-0.25$ day$^{-1}$.

Osteoclasts are multinucleated cells responsible for bone resorption, a process that is
accomplished in a fairly short time, relative to that required for bone formation. In cancellous bone, $30-40$ days versus $150$ days, \cite{Eriksen2010}. They originate from the hematopoietic monocyte-macrophage lineage. Osteoclastogenesis starts upon the release of RANKL and macrophage colony stimulating factor (M-CSF), \cite{Collin-Osdoby2001}. Osteoclasts precursors are monocytic cells that reside in the bone marrow. Its fusion into multinucleated, mature osteoclasts can only occur at the bone surface and must be completed for
successful resorption, \cite{Soe2020}. The number of nuclei per cell determines osteoclasts
resorption activity, \cite{Moller2020}. Overproduction of RANKL, as in the dysplastic tissue, results in the ectopic formation of numerous, large osteoclasts that excessively erode healthy bone, \cite{Riminucci2003,Whitlock2023}. Therefore,  the parameter that measures the resorption activity of osteoclast, $\delta$, has a lower value in healthy ($\delta_h$) than in dysplastic bone ($\delta_d$). The same applies for $\tau_c$, which measures osteoclasts activation. As for $\gamma_c$, traditionally, osteoclasts have been thought to be terminally differentiated cells that undergo apoptosis after a short lifespan of two weeks, \cite{Manolagas2000}. However, this has been
questioned by a recent study that showed osteoclasts with a lifespan of around six months, \cite{Jacome-Galarza2019}. This long-lived osteoclasts experiment iterative fusion with circulating blood monocytic cells. Recent research has shown that apoptosis is very rare in osteoclasts. Instead, osteoclasts recycle by fissioning into smaller more motile cells, called osteomorphs, and then fusing to form osteoclasts in a different location, \cite{McDonald2021}. Therefore, the value of the parameter $\gamma_c$ is just the inverse of the resorption period in trabecular bone, i.e., $\gamma_c\in[1/40-1/30]$, instead of the inverse of osteoclasts lifespan.

The differentiation of SSCs into oteoprogenitors is regulated by $\mu$. It is known that RANKL binding to RANK expressed in SSCs activates RANKL forward signaling, which inhibits osteoblast differentiation, \cite{Chen2018}. Therefore, $\mu$ cannot be larger than $1$, because the term regulating the differentiation flow loses its biological meaning when it becomes negative.

Experimental data have confirmed that the proportion
of mutant SSCs changes substantially with age. FD lesions in young individuals contain an expanded pool of SSCs compared with age-matched normal subjects, and a high proportion of them carry the mutation, whereas the proportion of mutant SSCs in older patients decreases drastically. This phenomenon is known as the age-dependent normalization of FD lesions, \cite{Kuznetsov2008}. Since we are working with young adults for which we are considering the SSC population to be constant, we get $\tau_m\geq\tau_s$. Notice that the population of SSCs is also expected to depend on the mutational load. 

As to the other parameters,  $\rho, \mu, \tau_s, \tau_{c,h}, \delta_h,$ and $\delta_m$, the strategy we have followed is estimating those of the healthy bone model so that the equilibrium point gives us a cell population distribution consistent with what it is found in the literature: osteocytes represent $90\%-95\%$ of
bone cells in the adult skeleton, \cite{Bonewald2011}, whereas osteoblasts constitute $4\%-6\%$ of all bone cells, \cite{Capulli2014}.

\begin{table}
\centering
%\begin{adjustbox}{width=1\textwidth}
      \begin{tabular}{ |c|c|c|c|c| } 
 \hline
Parameter & Meaning & Value & Units & Source\\  
 \hline
  &  &  &  & \\ 
$\rho$ & Osteoprogenitor proliferation  & $0.135 - 0.165$ & day$^{-1}$ & Estimated from\\ 
 & rate &  &  &  \cite{Bonewald2011,Capulli2014,Bolamperti2022} \\ 
    &  &  &  & \\ 
 \hline
  &  &  &  & \\ 
 $\rho_m$ & Mutant  osteoprogenitor  &  $2\rho-3\rho$ & day$^{-1}$ & \cite{Marie1997}\\ 
 &  proliferation rate &  &  & \\ 

 \hline
  &  &  &  & \\ 
$\alpha$& Differentiation rate &  $0.2-0.25$ & day$^{-1}$ & Estimated from\\ 
& & & &\cite{Bolamperti2022,Weng2022} \\ 
 &  &  &  & \\ 
 \hline
  &  &  &  & \\ 
$\mu$ & Saturation parameter in the & $<1$ & dimensionless &  Estimated from \\ 
 &  differentiation of SSCs. &  &  & \cite{Bonewald2011,Capulli2014,Bolamperti2022} \\ 
  &   &  &  & \\ 
 \hline
  &  &  &  & \\ 
$\tau_s$ & Incoming flow of osteoprogenitor  & $0.18 - 0.22$ & day$^{-1}$ & Estimated from \\
 & cells from SSCs. &  &  & \cite{Bonewald2011,Capulli2014,Bolamperti2022} \\ 
  &  &  &  & \\ 
 \hline
  &  &  &  & \\ 
  $\tau_{c,h}$ & Formation and activation of     & $(2- 8)\times 10^{-6}$ & day$^{-1}$ & Estimated from \\
  & osteoclasts due to the signaling  &  &  & \cite{Bonewald2011,Capulli2014,Bolamperti2022} \\ 
    &  from osteoprogenitors and   &  &  & \\ 
    &  osteocytes in healthy bone. &  &  & \\ 
      &  &  &  & \\ 
   \hline
    &  &  &  & \\ 
  $\tau_{c,d}$ & Formation and activation of    & $\tau_{c.d}\geq\tau_{c,h}$ & day$^{-1}$ & \cite{Whitlock2023}\\
  & osteoclasts due to the signaling &  &  & \\ 
    &  from osteoprogenitors and  &  &  & \\ 
      & osteocytes in dysplasic bone. &  &  & \\ 
          &  &  &  & \\ 
 \hline
     &  &  &  & \\ 
 $\gamma_c$ & Inverse of resorption time  & $1/40-1/30$ & day$^{-1}$ & \cite{Eriksen2010}\\
  & in cancellous bone. &  &  & \\ 
      &  &  &  & \\ 
   \hline
       &  &  &  & \\ 
$\delta_h$ & Efficiency dissolving the bone by & $90-120$ & day$^{-1}$ & Estimated from \\
  & osteoclasts (healthy bone). &  &  & \cite{Bonewald2011,Capulli2014,Bolamperti2022} \\
      &  &  &  & \\ 
  \hline
      &  &  &  & \\ 
$\delta_d$ & Efficiency dissolving bone & $\delta_d>\delta_h$ & day$^{-1}$ & \cite{Whitlock2023}\\
& by osteoclasts (dysplastic bone). &  &  & \\
    &  &  &  & \\ 
\hline
      &  &  &  & \\ 
$\delta_m$ & Induction rate of WT   & $1 - 3$ & day$^{-1}$ &   Estimated from \\ 
 & phenocopying by mutant cells. &  &  & \cite{Bonewald2011,Capulli2014,Bolamperti2022} \\ 
       &  &  &  & \\ 
 \hline
       &  &  &  & \\ 
 $\tau_m$ & Incoming flow of    & $\tau_m \geqslant \tau_s$ & day$^{-1}$ & \cite{Kuznetsov2008}\\
  & osteoprogenitors from &  &  & \\  
         &  mutant  skeletal stem cells&  &  & \\ 
                &    &  &  & \\ 
 \hline
\end{tabular} 
      \caption{List of values of the parameters used in the article. For details on how the parameters were estimated, see Section \ref{Sec:estimation}.}
      \label{tab:parameters}
\end{table}

\section{Results}\label{Sec:results}

\subsection{Basic properties of the FD mathematical model \eqref{FD_simple_model}}

\begin{proposition}\label{FDmodel_theoretical_results}
For any positive  
initial data $(P(0),P_m(0), P_p(0), C_I(0), C_L(0))$, 
and all
parameters of the model being positive, there exists a 
non-negative solution to Eqs. \eqref{FD_simple_model}   
with domain $t \in [0,T_0]$, for some $T_0>0$, or $\mathbb{R}_0^+$, which is bounded, and it is the unique maximal non-negative solution with initial data $(P(0),P_m(0), P_p(0), C_I(0), C_L(0))$.
\end{proposition}
\begin{proof}

Let us first prove the boundedness of the solutions. To do so we assume that $(P(t), P_m(t), P_p(t), C_I(t), C_L(t))$ is a non-negative solution to Eqs. \eqref{FD_simple_model}.
From \eqref{FD_simple_model1}, we have that
\begin{multline}
 \cfrac{dP}{dt}=  -\rho P^2 +P \left[\rho (1- P_m-P_p-C_I)-\right. \\ \left.\alpha - \delta_m P_m -\tau_s\mu\right] + \tau_s(1-\mu(P_m+P_p+C_I)).
 \end{multline}
We can see that for each $t>0$, the right-hand side of the equation is a quadratic function of $P$, i.e.
$y(P) = aP^2+bP+c$, with $a=-\rho < 0$, $b=b(t)$, and $c=c(t)$.

If we prove that $\left\{P(t)/\frac{dP}{dt}(t)> 0\right\}$ is bounded, 
then 

\begin{equation}
P(t)\leq \mathrm{max}\left\{P(0)\right\}\bigcup\overline{\left\{P(t)/\frac{dP}{dt}(t)> 0\right\}},
\end{equation}
will follow. 
For each $t_0>0$  such that 
$\tfrac{dP}{dt}(t_0)>0$, let $r_1(t_0)$ and $r_2(t_0)$ denote the two real roots of $y(P)$ at $t_0$, with $r_1\leq r_2$. Then 
$$P(t_0)<r_2(t_0)=\dfrac{b(t_0)+\sqrt{b(t_0)^2+4 \rho c(t_0)}}{2 \rho}.$$
Both $b$ and $c$ are bounded from above. Let us prove that $b$ is also bounded from below, and the boudedness of $P$ will follow. Since $a$ is negative, $P(t_0)\geq 0$, and $\tfrac{dP}{dt}(t_0)>0$, either $b(t_0)\geq0$, or $b(t_0)<0$ and $c(t_0)=-\rho r_1(t_0) r_2(t_0)> 0$, see Fig. \ref{fig:Parabolas}. In the second case, from $c(t_0)>0$ we get 
\begin{equation}\label{ineq}
\dfrac{1}{\mu}>P_m(t_0)+P_p(t_0)+C_I(t_0)\geq P_m(t_0).
\end{equation}
Using inequality \eqref{ineq}, we can bound $b$ from below.

\begin{figure}[h]
\centering
\includegraphics[width=1.0\textwidth]{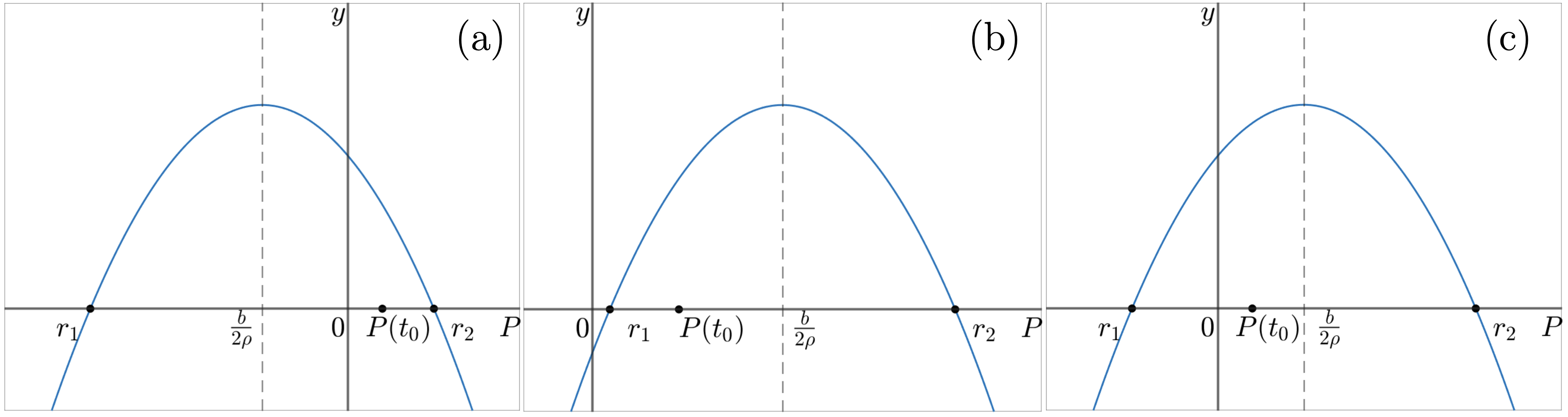}
\caption{Drawings of the quadratic function $y(P) =-\rho P^2+b P + c$ for (a) $b<0$, (b) $b\geq 0$ and $c>0$, and (c) $b>0$ and $c\leq 0$.}  
\label{fig:Parabolas}
\end{figure}

The boundedness of $P_m$ can be proved in an analogous way, using Eq. \eqref{FD_simple_model2}. 

As for $P_p$, from  Eq. \eqref{FD_simple_model3} we can write

\begin{equation}\label{Ppm}
\cfrac{dP_p}{dt}=  -\rho_m P_p^2 +P_p \rho_m (1- P-P_m-C_I)+\delta_m P P_m.
\end{equation}

For each $t>0$, the right-hand side of \eqref{Ppm} is again a quadratic function of the form,
$aP_p^2+bP_p+c$, with $a=-\rho_m$ negative and $c=\delta_m P P_m$ non-negative and bounded. Let us denote by $r_1(t)$ and $r_2(t)$ the smallest root and the largest root of the parabola at $t$, respectively. We do have $$0\leq P(t)<r_2(t)=\dfrac{b(t)+\sqrt{b(t)^2+4 \rho c(t)}}{2 \rho}.$$ Since $b$ is bounded from above and $c$ is bounded, it only remains to study the case in which $b(t)<0$. Let us proceed by reductio ad absurdum. If $\left\{P_p(t)/\frac{dP_p}{dt}(t)> 0, b(t)<0\right\}$ is not bounded, we can pick a sequence $\{t_n\}$ such that $\{P_p(t_n)\}\rightarrow +\infty$, and so, $\{r_2(t_n)\}\rightarrow +\infty$. Since $0<c=-\rho_m r_1 r_2\leq C$ for some constant $C$, $\{r_1(t_n)\} \rightarrow 0$. From that we would get that $\{b(t_n)\}\rightarrow 0$. Finally,

$$r_2(t_n)=\dfrac{b(t_n)+\sqrt{b(t_n)^2+4 \rho c(t_n)}}{2 \rho}\leq \dfrac{b(t_n)+\sqrt{b(t_n)^2+4 \rho C}}{2 \rho}\rightarrow \dfrac{\sqrt{\rho C}}{\rho},$$ 

which is a contradiction.

In order to show the boundedness of $C_I(t)$, we  prove that $M(t)$ is bounded. Indeed, if we add up Eqs.  \eqref{FD_simple_model1} to \eqref{FD_simple_model4}, we get 
\begin{multline*}
    \frac{dM}{dt} = -M \left(\rho P+\rho_m P_m+\rho_m P_p+\mu(\tau_m+\tau_s)\right)+ \\ \rho P+\rho_m P_m+\rho_m P_p+\tau_m+\tau_s-\delta C_I C_L.
\end{multline*}
Fix $t=t_0$ such that $\tfrac{dM}{dt}(t_0)>0$. 
Then 
\begin{multline*}
M(t_0)<\dfrac{\rho P(t_0)+\rho_m P_m(t_0)+\rho_m P_p(t_0)+\tau_m+\tau_s-\delta C_I(t_0) C_L(t_0)}{\rho P(t_0)+\rho_m P_m(t_0)+\rho_m P_p(t_0)+\mu (\tau_s+\tau_m)}\leq \\
\leq \dfrac{\rho P(t_0)+\rho_m P_m(t_0)+\rho_m P_p(t_0)+\tau_m+\tau_s}{\mu(\tau_m+\tau_s)}.
\end{multline*}
Since $P,P_m$ and $P_p$ are bounded, so does $M$ because $M(t)\leq \mathrm{max}\left\{M(0)\right\}\cup\overline{\left\{M(t)/\frac{dM}{dt}(t)> 0\right\}}$. As a consequence, $C_I$ is also bounded.

To prove the boundedness of $C_L$,  we proceed from \eqref{FD_simple_model5}. If $\tfrac{dC_L}{dt}(t_0)>0$ for some $t_0> 0$, then $$C_L(t_0)< \frac{\tau_C M(t_0)}{\gamma_c}.$$ Since $M$ is bounded, we can conclude as before.

Finally, the right-hand side of Eqs. \eqref{FD_simple_model} has continuous partial derivatives, and so the local existence and uniqueness are guaranteed. Thanks to the boundedness of non-negative solutions, we can restrict the domain of the right-hand side of the system to a compact set of $\mathbb{R}^5_{+,0}$. Therefore, our solution with initial data $(P(0),P_m(0),P_p(0),C_I(0),C_L(0))$ can be continued forward in time either infinitely or till the boundary of the compact set. The boundedness of the solution assures us that it can be continued forward infinitely, except if any of its components becomes $0$, where we can not guarantee its continuation as a non-positive solution. 
\end{proof}

\begin{proposition}\label{FDmodel_theoretical_results_positivity}
Assume that for some positive initial data $(P(0),P_m(0), P_p(0), C_I(0), C_L(0))$, 
and all the parameters of the model being positive, there exists a solution to Eqs. \eqref{FD_simple_model} %of the FD model  
either for $t>0$ or $0<t\leq T_0$ (for some positive $T_0$), such that $1-\mu M > 0$. % either for all $t>0$, or for $0<t\leq T_0$, for some positive $T_0$. 
Then, this solution is positive and bounded. Moreover, it can be continued forward to obtain the unique maximal non-negative solution to Eqs. \eqref{FD_simple_model}  %of the FD model 
with initial data $(P(0),P_m(0), P_p(0), C_I(0), C_L(0))$. 
\end{proposition}
\begin{proof}
%First, we prove the non-negativity of any solution. 
Consider a solution to Eqs. \eqref{FD_simple_model} with $1-\mu M > 0$ and non-negative initial data satisfying $P(0)P_m(0)>0$, and take the curve in $\mathbb{R}^5$ given by that solution,
$$X(t)=(P(t),P_m(t),P_p(t),C_I(t),C_L(t)).$$
Also, let $n_1=(-1,0,0,0,0), \ldots, n_5=(0,0,0,0,-1)$ denote the downward normal unit vectors to the hyperplanes $P=0, P_m=0, P_p=0, C_I=0,$ and $C_L=0$, respectively.  

Then, at the hyperplane $P=0$, $$\frac{dX}{dt}\cdot n_1 = -\tau_s (1-\mu P_m-\mu P_p-\mu C_I)=-\tau_s (1-\mu M)<  0.$$ 
Analogously, at the hyperplane $P_m=0$, 
$$\dfrac{dX}{dt}\cdot n_2 = -\tau_m (1-\mu P-\mu P_p-\mu C_I)=-\tau_m (1-\mu M)<  0 .$$ Hence, the curve $X(t)$ cannot intersect any of those hyperplanes. And so, both $P$ and $P_m$ are always positive. 

The non-negativity of the functions $P_p$ and $C_I$ follows from the non-negativity of $P$ and $P_m$. Indeed, at the hyperplane $P_p=0$, we do have $$\frac{dX}{dt} \cdot n_3 = -\delta_m P P_m< 0.$$  
Whereas at the hyperplane $C_I=0$, $$\frac{dX}{dt} \cdot n_4 = -\alpha P<  0.$$ Therefore, both $P_p$ and $C_I$ are always positive, except at $t=0$, where they are non-negative. 

For the last function, at the hyperplane $C_L=0$, we do have $$\frac{dX}{dt} \cdot n_5 = -\tau_c(P+P_m+P_p+C_I)< 0,$$ so $C_L$ is non-negative and it can only vanish at $t=0$.

Once the non-negativity of the solutions is proved, Proposition \ref{FDmodel_theoretical_results} applies, and so boundedness follow. The continuation and uniqueness is proved following the arguments in the final part of the proof of Proposition \ref{FDmodel_theoretical_results}.
\end{proof}

\begin{remark}\label{Remark:-solFD} Solutions  to Eqs. \eqref{FD_simple_model}  taking negative values do exist. For instance, take $\varepsilon>0$ and consider the (local) unique solution around $t=0$ with Cauchy data $(0,P_m(0),P_p(0),\dfrac{1}{\mu}+\varepsilon,C_L(0))$. Since $\dfrac{dP}{dt}(0)<0$ and $P(0)=0$, $P$ will take negative values for positive values of $t$ close enough to $0$. 

Therefore, the hypothesis $1-\mu M>0$ is not only reasonable from a biological point of view (in order to have a positive flow from SSCs), but it is also needed from a mathematical point of view to prove positiveness. 

\end{remark}

\subsection{Properties of the healthy bone model}
\begin{proposition}\label{HB_model_theoretical_results}
For any positive initial data $(P(0), C_I(0), C_L(0))$,
and all
parameters of the model being positive, there exists a
non-negative solution to Eqs.  \eqref{HB_simple_model} 
with domain $[0,T_0]$, for some $T_0>0$, or $\mathbb{R}_0^+$, which is bounded, and it is the unique maximal non-negative solution with initial data $(P(0), C_I(0), C_L(0))$.
\end{proposition}
\begin{proof}
The proof mimics the steps of the proof of Proposition \ref{FDmodel_theoretical_results}, so the
details are omitted.
\end{proof}

\begin{proposition}\label{HB_model_theoretical_results_positivity}
Assume that for some positive initial data $(P(0), C_I(0), C_L(0))$
and all the parameters of the model being positive, there exists a solution to equations \eqref{HB_simple_model1} --\eqref{HB_simple_model4} 
either for $t>0$ or $0<t\leq T_0$, for some positive $T_0$, such that such that $1-\mu M > 0 $.
Then, this solution is positive and bounded. Moreover, it can be continued forward to obtain the unique maximal non-negative solution to equations \eqref{HB_simple_model1} --\eqref{HB_simple_model4}
with initial data $(P(0), C_I(0), C_L(0))$. 
\end{proposition}

\begin{proof}
Analogous to the proof of Proposition \ref{FDmodel_theoretical_results_positivity}.
\end{proof}

\begin{remark} Solutions to the Eqs. \eqref{HB_simple_model} taking negative values do exist. Take $\varepsilon>0$ and consider the (local) unique solution with Cauchy data $\left(0,\dfrac{1}{\mu}+\varepsilon,C_L(0)\right)$. Since $\tfrac{dP}{dt}(0)<0$ and $P(0)=0$, $P$ will take negative values for positive values of $t$ close enough to $0$.

As discussed in the previous remark, the hypothesis $1-\mu M>0$ is justified both from a biological and a mathematical point of view.
\end{remark}

\begin{proposition}
\label{HB_model_theoretical_equilibrium}
For any positive choice of the parameters of the model with $\rho<\alpha$, 
Eqs. \eqref{HB_simple_model} have a unique non-negative equilibrium point that it is uniformly asymptotically stable. 
\end{proposition}
\begin{proof}
For the equilibrium point, the following conditions must hold
\begin{subequations}\label{HB_simple_model_equilibrium}
\begin{eqnarray}
    0 &= & \rho P (1-P-C_I) + \tau_s (1-\mu P -\mu C_I)) - \alpha P,\\
    0& = & \alpha P - \delta C_I \cdot C_L,\\
    0 &= &  \tau_c (P+C_I) - \gamma_c C_L.
\end{eqnarray}
\end{subequations}
From Eqs. (\ref{HB_simple_model_equilibrium}) we obtain the relations $P= \frac{\delta C_I \cdot C_L}{\alpha}$ and $P+C_I = \frac{\gamma_c C_L}{\tau_c}$, that allow us to rewrite the system in the following equivalent form
\begin{subequations}\label{HB_sme}
\begin{eqnarray}
    P &= & \frac{\delta \gamma_c C_L^2}{\tau_c(\alpha+\delta C_L)},\\
    C_I &= & \frac{\alpha \gamma_c C_L}{\tau_c(\alpha+\delta C_L)},\\
    0 &= &  -\frac{\rho\delta \gamma_c^2}{\tau_c} {C_L}^3+\delta \gamma_c(\rho - \alpha - \mu\tau_s ){C_L}^2+\tau_s(\tau_c\delta - \mu\alpha\gamma_c){C_L}+\alpha\tau_s\tau_c.
\end{eqnarray}
\end{subequations}
Since we are  interested only on non-negative solutions, the previous expressions for $P$ and $C_I$ always make sense. 

Consider the function $f(x) = a_3 x^3 + a_2 x^2+ a_1 x + a_0$, where $a_3 = -\rho\delta \gamma_c^2/\tau_c$, $a_2 = \delta \gamma_c(\rho - \alpha - \mu\tau_s )$, $a_1=\tau_s(\tau_c\delta - \mu\alpha\gamma_c)$, and $a_0=\alpha\tau_s\tau_c$. As $f(0) = a_0$ is positive and $\displaystyle\lim_{x\to \infty}f(x)=-\infty$, there exists at least one positive root of the equation $f(x)=0$. 
At the same time, by Descartes' rule of signs, the number of positive roots is either equal or less than one,  as $a_3<0$, $a_2<0$, and $a_0>0$. Hence,  the last equation of the system has a unique non-negative solution, and therefore, Eqs. \eqref{HB_simple_model} have a unique non-negative equilibrium point.

In order to prove the  stability, consider the Jacobian matrix
\begin{equation}
J = \begin{pmatrix}\label{HB_simple_model_par_deriv}
-\rho P + \rho(1-M)-\mu \tau_s - \alpha & \quad -\rho P-\mu \tau_s& \quad 0\\
\alpha  &\quad - \delta C_L&\quad - \delta C_I\\
\tau_c &\quad \tau_c &\quad   - \gamma_c,\\
\end{pmatrix},
\end{equation}
and its characteristic equation 
\begin{equation}
P(\lambda) = \lambda^3+b_2 \lambda^2 + b_1 \lambda +b_0,
\end{equation}
 where $b_2=\rho P + \mu \tau_s + \alpha - \rho(1-M)+\delta C_L+ \gamma_c$, $b_1= (\delta C_L + \gamma_c) (\rho P + \mu \tau_s + \alpha - \rho (1-M))+\alpha (\rho P + \mu \tau_s)+\delta(\gamma_c C_L+\tau_c C_I)$, and $b_0=(\rho P + \mu \tau_s + \alpha - \rho(1-M))(\delta\gamma_c C_L+\delta\tau_c C_I))+(\alpha\gamma_c - \delta\tau_c C_I)(\rho P + \mu \tau_s)$.

Then, by Routh-Hurwitz criterion, $P(\lambda)$ has all roots in the open left half-plane, since $b_2, b_1$ and $b_0$ are positive (because of inequality $\alpha>\rho$) and $b_2b_1>b_0$. Therefore, by a classical ODE theorem \cite{Lukes}, the equilibrium point is uniformly asymptotically stable.
\end{proof}

\subsection{Numerical Simulations}\label{results}

\subsubsection{Mathematical model for healthy bone provides the value of some parameters needed for the dysplastic bone model}

Figure \ref{fig:simul_Bone_model}(a) shows a typical example of the dynamics of the healthy bone described by Eqs.  \eqref{HB_simple_model}. As it has been previously mentioned, there are many mathematical models of healthy bone, and our simplified approach here is developed to aid in the parameter fitting tasks and to provide a ground on which to build the FD model.  There are several parameters related to the healthy bone behavior whose values are difficult to estimate, but turn out to be key in our model of dysplastic bone. A reasonable value for those parameters have been chosen so that in the equilibrium, the proportions of cells in the different compartments match their real-world abundances. More specifically, osteocytes are known to be around $90\%$ of the bone in young adults and osteoprogenitors around $5\%$.  This is consistent with the fact that osteocytes represent $90\%-95\%$ of bone cells in the adult skeleton, \cite{Bonewald2011}, whereas osteoblasts constitute $4\%-6\%$ of all bone cells, \cite{Capulli2014}. During the first three months of the simulation, osteocytes expanded, showing a peak, and then their numbers stabilised. Osteoprogenitor cells expanded during the first two weeks of the simulation, till showing a peak. Afterward, their numbers stabilized and began to decrease, eventually reaching a value around $5\%$ of the total bone. Notice that the equilibrium point does not depend on the initial condition chosen, but only on the value of parameters (Proposition \ref{HB_model_theoretical_equilibrium}).

\subsubsection{Mathematical model for dysplastic bone shows that even if the departure percentage of mutant cells is small, mutant and WT phenocopying osteoprogenitors become the main populations}

For the initial conditions of the FD model, we haven chosen the values given by the equilibrium point of the healthy bone model, and we have added $P_m(0)=10^{-4}$ and $P_p(0)=0$. Let us remark that no estimation can be found in the literature of the initial mutational load of a bone region affected by FD. It is thought to depend on the moment at which the initial mutation took place, which is different in each patient, and impossible to be determined. The mutation that gives rise to FD happens in a single cell during embryonic development, \cite{Zhao2018}, either after or before gastrulation. Even if it is not known if the mutation represents an evolutionary advantage or disadvantage for the mutant initial cell in comparison with the WT ones, \cite{Hartley2019}, the percentage of mutant cells before lesions become visible is expected to be very small.

\begin{figure}
\centering
\includegraphics[width=\textwidth]{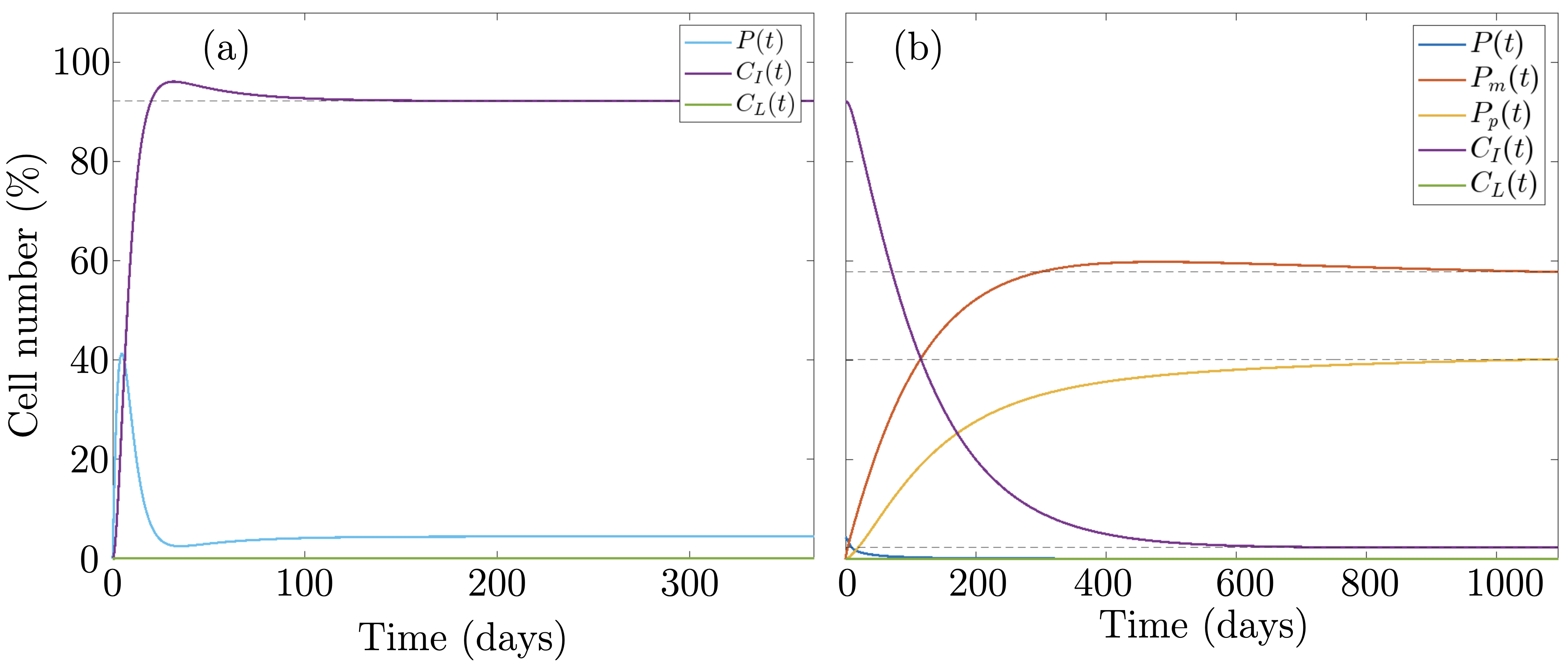}
    \caption{(a) Time dynamics of osteoprogenitor cells (blue), osteocytes (purple), and  osteoclasts (green) compartments according
to Eqs. \eqref{HB_simple_model} for parameters $\rho = 0.15$ day$^{-1}$, $\alpha = 0.2$ day$^{-1}$, $\mu = 0.99$, $\tau_s = 0.2$ day$^{-1}$,$\tau_c = 1/350000$ day$^{-1}$, $\gamma_c = 1/35$ day$^{-1}$, and $\delta=90$ day$^{-1}$.  (b) Dynamics of osteoprogenitor cells (blue), mutant osteoprogenitors
(red), WT phenocopying osteoprogenitors (yellow), osteocytes (violet), and osteoclasts (green) compartments according to Eqs. \eqref{FD_simple_model} for parameters $\rho = 0.15$ day$^{-1}$, $\rho_m = 0.3$ day$^{-1}$, $\alpha = 0.2$ day$^{-1}$, $\mu = 0.99$, $\tau_s = 0.2$ day$^{-1}$,$\tau_c = 1/350000$ day$^{-1}$, $\gamma_c = 1/35$ day$^{-1}$, $\delta=100 $ day$^{-1}$, $\delta_m=1$ day$^{-1}$, $\tau_m=0.2 $ day$^{-1}$.   \label{fig:simul_Bone_model}} 
\end{figure}

Figure \ref{fig:simul_Bone_model}(b), shows an example of the typical dynamics of the disease burden in a dysplastic bone ruled by Eqs. \eqref{FD_simple_model}, where mutant osteoprogenitors are present. We can observe that mutant and WT phenocopying osteoprogenitors expanded, while osteocytes experience a continuous decrease, as it was pointed out in \cite{Marie1997,Riminucci1997}. Their numbers stabilized after approximately $600$ days. Even for very small values of $P_m(0)$, in the equilibrium, most cells are immature progenitors and either bear the mutation or are WT phenocopying cells, in agreement with experimental observations \cite{Kuznetsov2008}.

\subsubsection{Parameters influencing the severity of the disease}\label{param-results}

The parameters that characterize the disease include: $\delta_m$, indicating the efficiency of mutant osteoprogenitors in converting WT osteoprogenitors into phenocopying WT ones; $\rho_m$, representing the actual proliferation rate of mutant osteoprogenitors; $\tau_m$, denoting the flow of cells from mutant SSCs; $\delta$, reflecting the efficiency of bone dissolution by osteoclasts, known to be higher in dysplastic bone compared to healthy bone; and $\tau_c$, signifying the formation and activation of osteoclasts due to signaling from osteoprogenitors and osteocytes, also observed to be higher in dysplastic bone than in healthy bone.

In fibrous dysplasia (FD), abnormal cell proliferation is acknowledged to occur at the osteoprogenitor level \cite{Marie1997}. However, given that RANKL plays a role in obstructing the flow from SSCs \cite{Chen2018}, and FD is marked by an increase in RANKL levels \cite{deCastro2019}, we have also taken into account the parameter $\mu$.

For each of these parameters, we have examined the variations in the equilibrium point as the parameter changes, keeping the other parameters constant. In all cases except for $\mu$, we operated under the assumption that larger values of these parameters correspond to more severe fibrous dysplasia (FD)—that is, a higher sum of the populations $P_m$ and $P_p$, and a smaller number of osteocytes at the equilibrium point.

\begin{figure}[H]\label{Equilibrium-point-and-parameters}
\centering
\includegraphics[width=\textwidth]{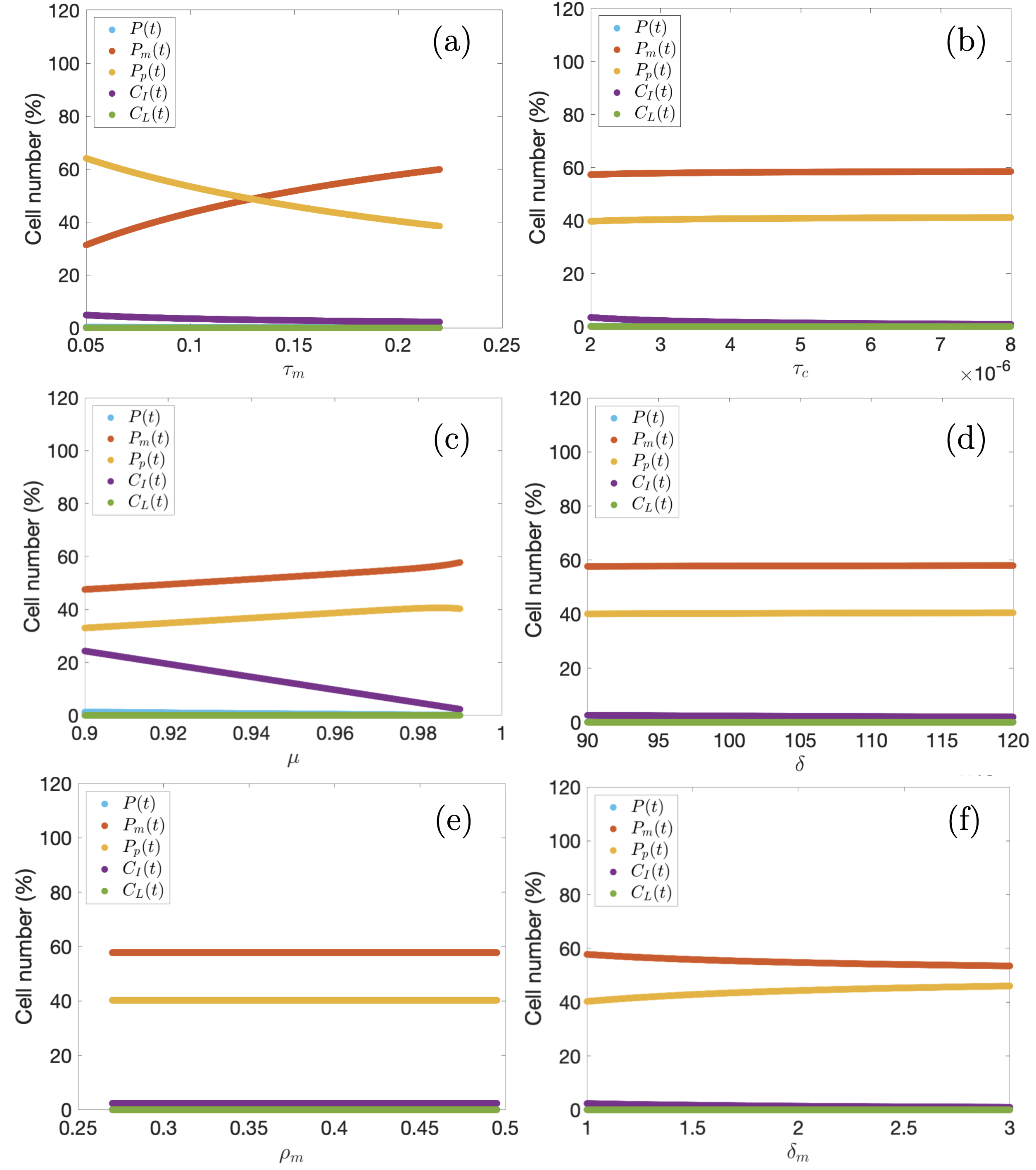}
    \caption{Dependence of the equilibrium point on the model parameter values, fixing all parameters except for one. The fixed parameters take the values $\rho = 0.15$ day$^{-1}$, $\rho_m = 0.3$ day$^{-1}$, $\alpha = 0.2$ day$^{-1}$, $\mu = 0.99$, $\tau_s = 0.2$ day$^{-1}$,$\tau_c = 1/350000$ day$^{-1}$, $\gamma_c = 1/35$ day$^{-1}$, $\delta=100 $ day$^{-1}$, $\delta_m=1$ day$^{-1}$, and $\tau_m=0.2 $ day$^{-1}$. Whereas the value of the varying parameter is specified in each case. (a) Varying parameter $\tau_m \in [0.05, 0.22]$, (b) Varying parameter $\tau_c \in [2 \cdot 10^{-6}, 8 \cdot 10^{-6}]$, (c) Varying parameter $\mu \in [0.9, 99]$, (d) Parameter values $\delta \in [90, 129]$, (e) Varying parameter $\rho_m \in [0,27, 0,495]$, (f) Varying parameter $\delta_m \in [1, 3]$.   \label{fig:Equilibrium_parameters}}  
\end{figure}

To our surprise, except for $\mu$, the changes experimented by $P_m+P_p$ and $C_I$ at the equilibrium point were very subtle, as it can be seen in Fig. \ref{fig:Equilibrium_parameters}. Whereas for $\delta_m$ and $\tau_m$ the values of $P_m$ and $P_p$ varied considerably, for $\tau_c$, $\delta$, and $\rho_m$ the variation was very small. As for $\mu$, it is the parameter for which the most significant changes in the equilibrium point can be observed: the closer to one, the more severe the total disease burden. In the concluding part of Section \ref{Sec:results}, considering all the preceding subsections, we conduct a detailed analysis of the parameters that either align with clinical observations or deviate from anticipated outcomes.

 \subsection{Sensitivity analysis}\label{subs_sensitivity}

Our equations \eqref{FD_simple_model}, encompass ten parameters: $\rho$, $\rho_m$, $\alpha$, $\mu$, $\tau_s$, $\tau_c$, $\gamma_c$, $\delta$, $\delta_m$, and $\tau_m$. To identify the model parameters with the most significant impact on the equilibrium point, we conducted a sensitivity analysis using Sobol's method \cite{Saltelli(2010)}. This method, also known as variance-based sensitivity analysis, operates within a probabilistic framework by decomposing the variance of the model's output into fractions attributable to inputs. Each sensitivity effect is expressed through a conditional variance and is computed by evaluating a multidimensional integral using a Monte Carlo method.

\begin{figure} 
\centering
\includegraphics[width=\textwidth]{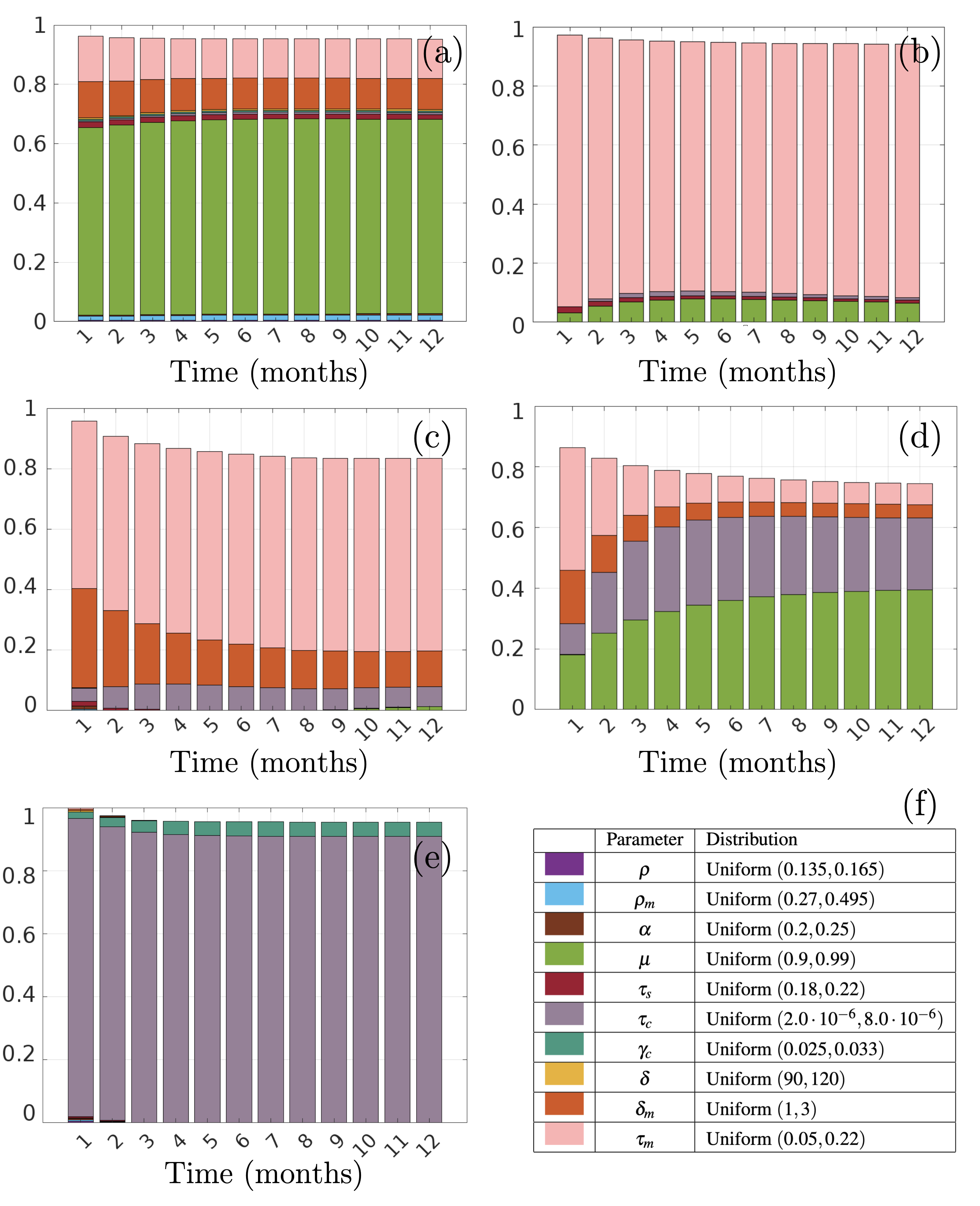} 
\caption{First order sensitivity indexes for: (a) $P(t)$, (b) $P_m(t)$, (c) $P_p(t)$\label{fig:sensetivity_P_p}
(d) $C_I(t)$, and (e) $C_L(t)$. Panel (f) indicates the color codes for each parameter and the range of values used in the sensitivity analysis.\label{fig:Sensetivity_Analusis}}
\end{figure}

We calculated the first-order sensitivity index in order to quantify the relative impact on the model output of the variability in a given parameter. The influence of that parameter on the model's response is indicated by the values of that index, so that the smallest the index, the smallest the impact. The first order index was calculated for the time points $30$, $60$, $91$, $121$, $152$, $182$, $213$, $244$, $274$, $305$, $335$ and $365$ days. Using a priori information on the parameters (Table \ref{tab:parameters}), we defined the uniform distribution functions. A set of $5000$ simulations (parameter values) was used to calculate the
sensitivity indices. 

Figure \ref{fig:Sensetivity_Analusis} summarizes the results of the sensitivity analysis of Eqs. \eqref{FD_simple_model}.
The parameters $\mu$, $\tau_m$, and $\tau_c$ exerted the most substantial influence on the solutions. Specifically, for the progenitor population $P(t)$, $\mu$ emerged as the paramount parameter, indicating that differentiation holds greater significance than proliferation for healthy osteoprogenitors. In the case of the mutant osteoprogenitor population $P_m(t)$, it is evident that $\tau_m$ assumes the highest level of importance. Furthermore, for the population of WT phenocopying osteoprogenitors $P_p(t)$, once again, $\tau_m$ played the most significant role, while $\delta_m$ also demonstrated importance, albeit diminishing over time.

The pivotal role of $\tau_m$ in both $P_m$ and $P_p$ aligns with the well-known age-dependent normalization of FD lesions. In this context, mutant SSCs cells fail to self-renew and undergo apoptosis, contrasting with WT SSCs that survive and enable the formation of structures that are histologically normal \cite{Kuznetsov2008}.

Concerning the mature osteocytes $C_I(t)$, the parameter $\tau_m$ emerged as the most crucial in the initial two months, underscoring the impact of the flow from SSCs on the osteocyte population. Subsequently, both $\mu$ and $\tau_c$ gained significance. This suggests that the growth in the number of osteoclasts has a more pronounced effect on osteocytes than the enhancement of their activity. Additionally, the increased competition for space in the bone also influences osteocytes.

For the population of osteoclasts $C_L(t)$, only $\tau_c$ played a pivotal role, aligning with the observed active osteoclastogenesis in FD lesions \cite{Riminucci2003}.

\subsection{Linear stability}\label{subs_stability}

In Proposition \ref{HB_model_theoretical_equilibrium}, we have proven that  Eqs. \eqref{HB_simple_model}  have a unique non-negative equilibrium point that is uniformly asymptotically stable. Because of the mathematical complexity of the Eqs. \eqref{FD_simple_model}, we cannot obtain similar results for the full FD model. However, the conditions for the equilibrium point
\begin{subequations}
\label{FD_equilibrium_point}
\begin{eqnarray}
    0 & = &  \rho P (1 -  M) + \tau_s (1-\mu M) - \alpha P - \delta_m P P_m,  %\label{FD_ep1}
    \\
    0 & = & \rho_m P_m (1 -  M) + \tau_m (1-\mu M), %\label{FD_ep2}
    \\
    0 & = & \rho_m P_p (1 -  M) + \delta_m P P_m, %\label{FD_ep3}
    \\
    0 & = &  \alpha P - \delta C_I C_L,%\label{FD_ep4}
    \\
    0  & = &  \tau_c M - \gamma_c C_L,%\label{FD_ep5}
    \\
    M  & =  &P+P_m+P_p+C_I, %\label{FD_simple_model6}%\\
    %L &=  & P+P_m+P_p+C_I+C_L \label{FD_simple_model7} 
    \end{eqnarray}
\end{subequations}
can be rewritten expressing each component in terms of $P_m$, as follows

\begin{subequations}
\begin{eqnarray}
      P & = & \frac{\tau_s\rho_m(1-\mu)P_m}{\rho_m \delta_m P_m^2 + P_m (\alpha\rho_m+\mu\delta_m \tau_m) +\tau_m(\alpha\mu+(1-\mu)\rho)},\\
     P_p & = & \frac{\tau_s \delta_m  P_m^2(\rho_m P_m +\mu\tau_m )}{\tau_m(\rho_m \delta_m P_m^2 + P_m (\alpha\rho_m+\mu\delta_m \tau_m) +\tau_m(\alpha\mu+(1-\mu)\rho))}, \\
       M  & = & \frac{\rho_m P_m+ \tau_m}{\rho_m P_m+ \mu\tau_m},  \\    
    C_I & = & \frac{\alpha\gamma_c\tau_s\rho_m(1-\mu)(\rho_m P_m+ \mu\tau_m)P_m/ \delta \tau_c}{\rho_m \delta_m P_m^2 + P_m (\alpha\rho_m+\mu\delta_m \tau_m) +\tau_m(\alpha\mu+(1-\mu)\rho)(\rho_m P_m+ \tau_m)  },\\
      C_L & = & \frac{\tau_c(\rho_m P_m+ \tau_m)}{\gamma_c(\rho_m P_m+ \mu\tau_m)},\\
       0 & = & c_5 P_m^5 +c_4 P_m^4 + c_3 P_m^3 +c_2 P_m^2 +c_1 P_m +c_0, \\     
\end{eqnarray}
\end{subequations}
where the coefficients $c_j$ are given by the expressions
\begin{eqnarray*}
c_0 &=& -\tau_m^3(\alpha\mu + \rho(1-\mu)), \\
c_1 &= & \tau_m^2((\mu\tau_m- 2\rho_m)(\alpha\mu + \rho(1-\mu)) - \alpha\rho_m - \delta_m\mu\tau_m + \mu\rho_m\tau_s(1-\mu) \\
& &+ \alpha\gamma_c\mu^2\rho_m\tau_s(1-\mu)/(\delta\tau_c)),\\ 
c_2 &= &\tau_m(\tau_m\alpha\mu \rho_m (1 +\mu) - \delta_m\rho_m\tau_m - 2\rho_m(\alpha\rho_m + \delta_m\mu\tau_m)-\rho_m^2(\alpha\mu + \rho(1-\mu))+ \\  & & \mu\tau_m(\alpha\rho_m + \delta_m\mu(\tau_m+\tau_s)) +(1-\mu)((\tau_m\rho\rho_m+\rho_m^2\tau_s)(1 +\mu) \\
& & +  2\alpha\gamma_c\mu\rho_m^2\tau_s/(\delta\tau_c))),\\
 c_3 &=&\rho_m
\left((\alpha\rho_m + \delta_m\mu\tau_m)(\tau_m(1+\mu)- \rho_m)  
+\delta_m\tau_m(2\mu\tau_s+ \mu\tau_m  + \mu^2\tau_s- 2\rho_m) \right.
\\ & & + \left.\rho_m^2\tau_s(1-\mu)+ \rho_m\tau_m(\alpha\mu + \rho(1-\mu))  + (\alpha\gamma_c\rho_m^2\tau_s(1-\mu))/(\delta\tau_c)\right)),\\ %\quad \text{ if } \rho_m<\mu(\tau_s+\tau_m/2) 
c_4 & =&  \rho_m^3 \alpha+\rho_m^2\delta_m((\tau_m+\tau_s)(1+2\mu)-\rho_m),\\% \quad \text{ if } \rho_m<(1+2\mu)(\tau_s+\tau_m) \\
 c_5 &=& \delta_m\rho_m^3 (1+\tau_s/\tau_m).
\end{eqnarray*}

Let us consider the polynomial given by $f(x) = c_5 x^5 +c_4 x^4 + c_3 x^3 +c_2 x^2 +c_1 x +c_0$. As $f(0) = c_0$ is negative and $\displaystyle\lim_{x\to \infty}f(x)=+\infty$, there exists at least one positive root of the equation $f(x)=0$.  At the same time, $c_0< 0$ and $c_5>0$. Given the complexity of the values of $c_1$, $c_2$, $c_3$, and $c_4$, we can impose different assumptions on the parameters such that there is only one change of sign in the parameters, meaning that, by Descartes' rule of signs, $f$ has only one positive root. And so, Eqs. \eqref{FD_simple_model} has a unique positive equilibrium point. That is the case in a neighborhood of our election of parameters, see Fig. \ref{fig:simul_Bone_model} (b). 

As for the stability, we have studied numerically broad ranges of the biologically relevant parameters obtained in the sensitivity analysis of Section \ref{subs_sensitivity}, $\mu$, $\tau_c$, and $\tau_m$. We have varied the parameter $\mu$ from $0.5$ to $1.1$,  $\tau_c$ from $2.5\cdot10^{-6}$  to $5\cdot10^{-5}$, and $\tau_m$ from $0.01$ to $0.5$, while maintaining the other parameters fixed according to Table \ref{tab:parameters}. We have observed that for $\mu<1$, we get uniformly asymptotic stability, whereas for $\mu>1$, instability arises, no matter the values of $\tau_c$ and $\tau_m$. Which tell us that $\mu<1$ is not only a necessary condition from a biological point of view, but also undesirable mathematical behaviours occur when this assumption is not considered.

\subsection{Analysis of the main parameters in the FD model}

\subsubsection{The behaviour of the flow of mutant osteoprogenitors differentiated from mutant SSCs supports age-dependent normalization of FD lesions}

\begin{figure}[H]
\centering
\includegraphics[width=0.8\textwidth]{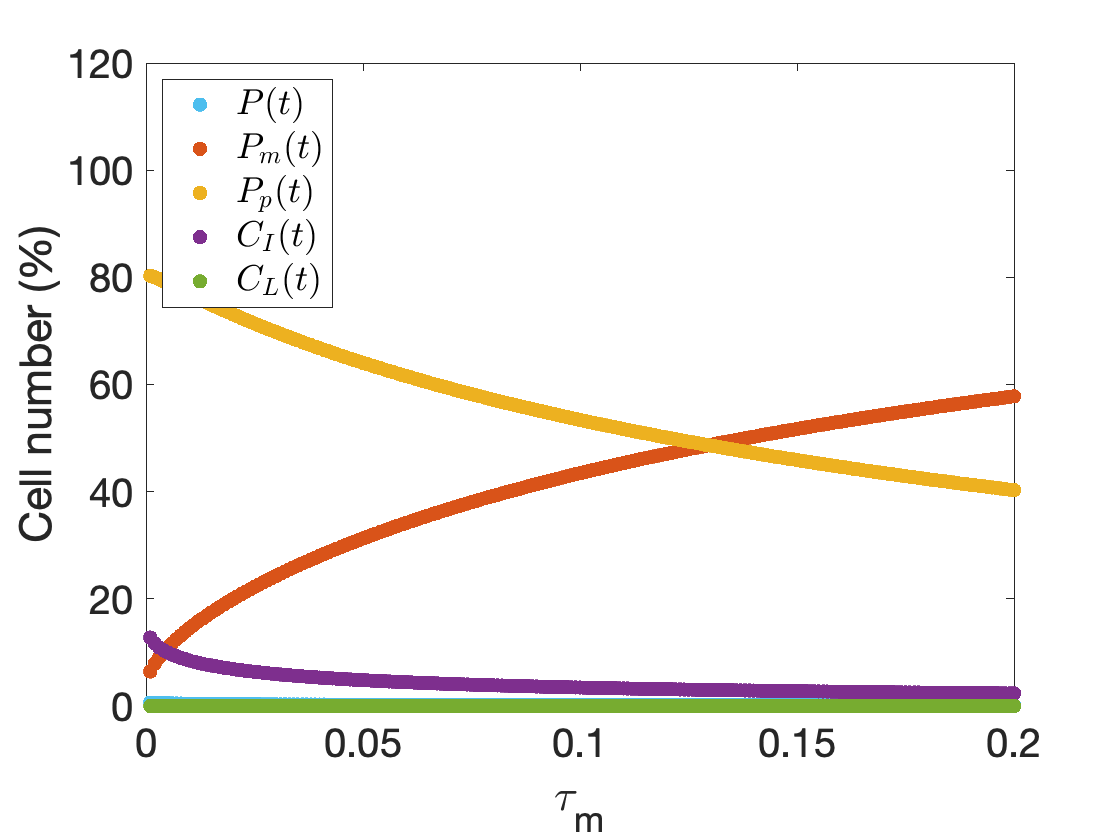}
    \caption{Dependence of the equilibrium point on the value of $\tau_m$ when it takes values close to $0$, fixing the other parameters: $\rho = 0.15$ day$^{-1}$, $\rho_m = 0.3$ day$^{-1}$, $\alpha = 0.2$ day$^{-1}$, $\mu = 0.99$, $\tau_s = 0.2$ day$^{-1}$,$\tau_c = 1/350000$ day$^{-1}$, $\gamma_c = 1/35$ day$^{-1}$, $\delta=100 $ day$^{-1}$, and $\delta_m=1$ day$^{-1}$.     \label{tau_m-0}}
\end{figure}

The parameter $\tau_m$ controls the incoming flow of mutant osteoprogenitors differentiating from mutant SSCs. In the sensitivity analysis, it is the most important parameter for both mutant osteoprogenitors and WT phenocopying osteoprogenitors. In the analysis of the equilibrium point in terms of $\tau_m$ shown in Figure \ref{tau_m-0}, the sum of both populations is almost constant and responsible for more than the $90\%$ of the bone cells. In the same figure, it can be checked that which the largest population is depends on the value of $\tau_m$.

The phenomenon known as age-dependent normalization of FD lesions, \cite{Kuznetsov2008}, consists in the drastic decrease of the proportion of mutant SSCs in older patients, due to the fact that mutant SSCs enter senescence before WT SSCs do, and their progeny is consumed by apoptosis.
This phenomenon cannot be observed in Figure \ref{tau_m-0}, because the choice of parameters for our numerical simulations has been made for a young adult. But if we consider values of $\tau_m$ close to $0$, as in Figure \ref{tau_m-0}, we can observe that the population of mutant osteoprogenitors almost disappear, whereas the population of osteocytes increases. The population of WT phenocopying osteoprogenitors does not disappear because in our model we have not included the reversibility of the phenotype, due to the fact that in young adults the lesions are stable and no such phenomenon is expected.

\subsubsection{The saturation parameter in the differentiation of mutant SSCs and the severity of FD}

In contrast with the parameter previously analysed, the saturation parameter that modulates the flow affects the differentiation of both WT and mutant SSCs. From a biological point of view, it was clear that $\mu<1$. Otherwise, the term measuring the increase in the population of osteoprogenitors (either WT or mutant) due to the differentiation of SSCs (WT or mutant, respectively) could be negative, which constitutes a biological nonsense. We have obtained that also from a mathematical point of view the parameter $\mu$ must be less than $1$. This is because for a reasonable choice of the rest of parameters, the condition $\mu< 1$ is needed to obtain a unique uniformly asymptotic stable equilibrium point.

\begin{figure}[H]
\centering
   \includegraphics[width=0.8\textwidth]{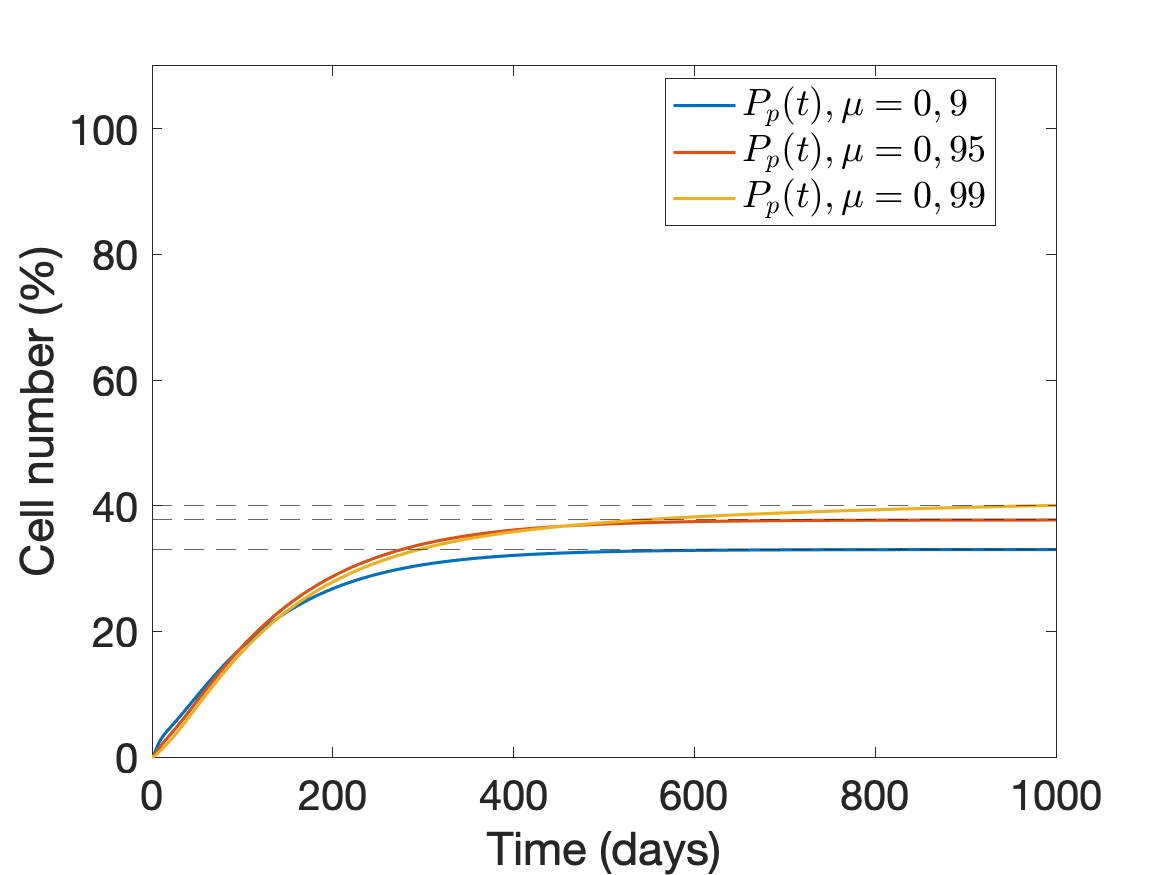} \\
	%\textbf{(a)}   \\[6pt]
	\includegraphics[width=0.8\textwidth]{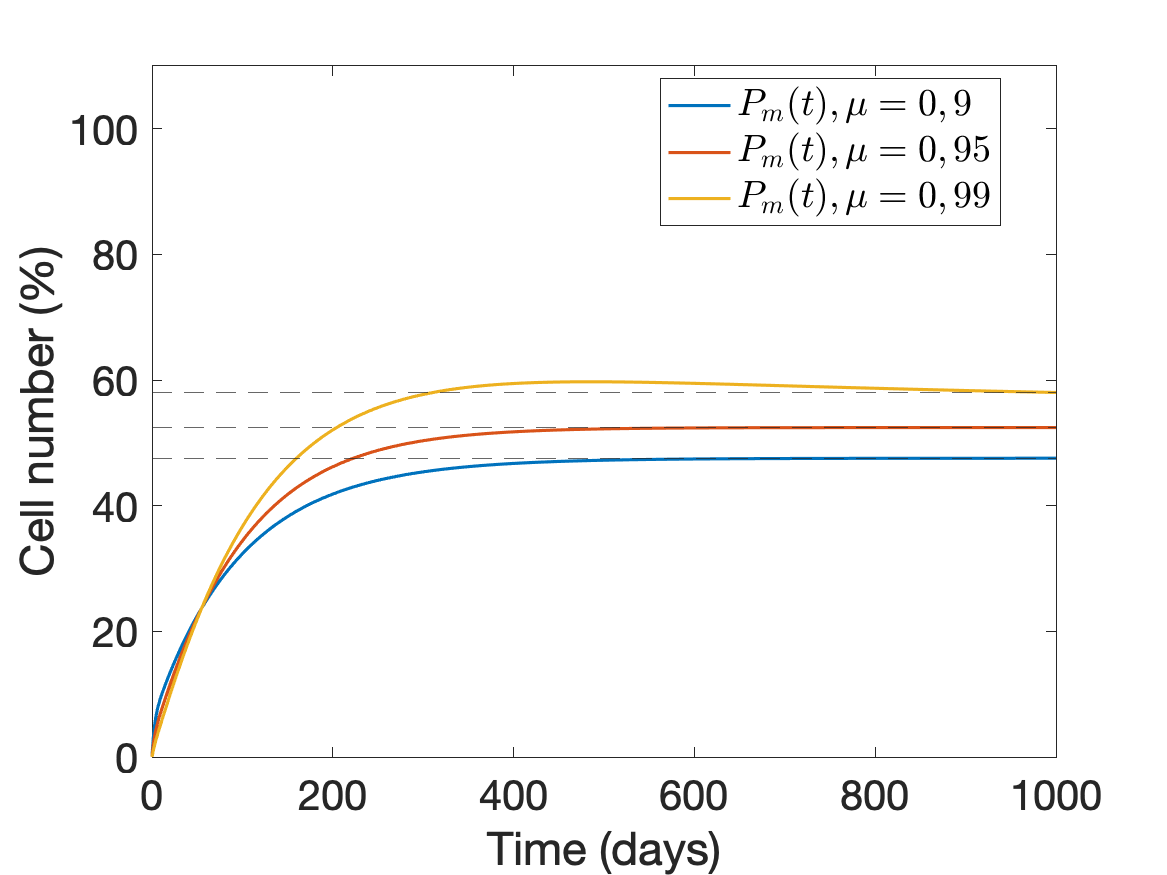} \\
	%\textbf{(b)} \\[6pt]
    \includegraphics[width=0.8\textwidth]{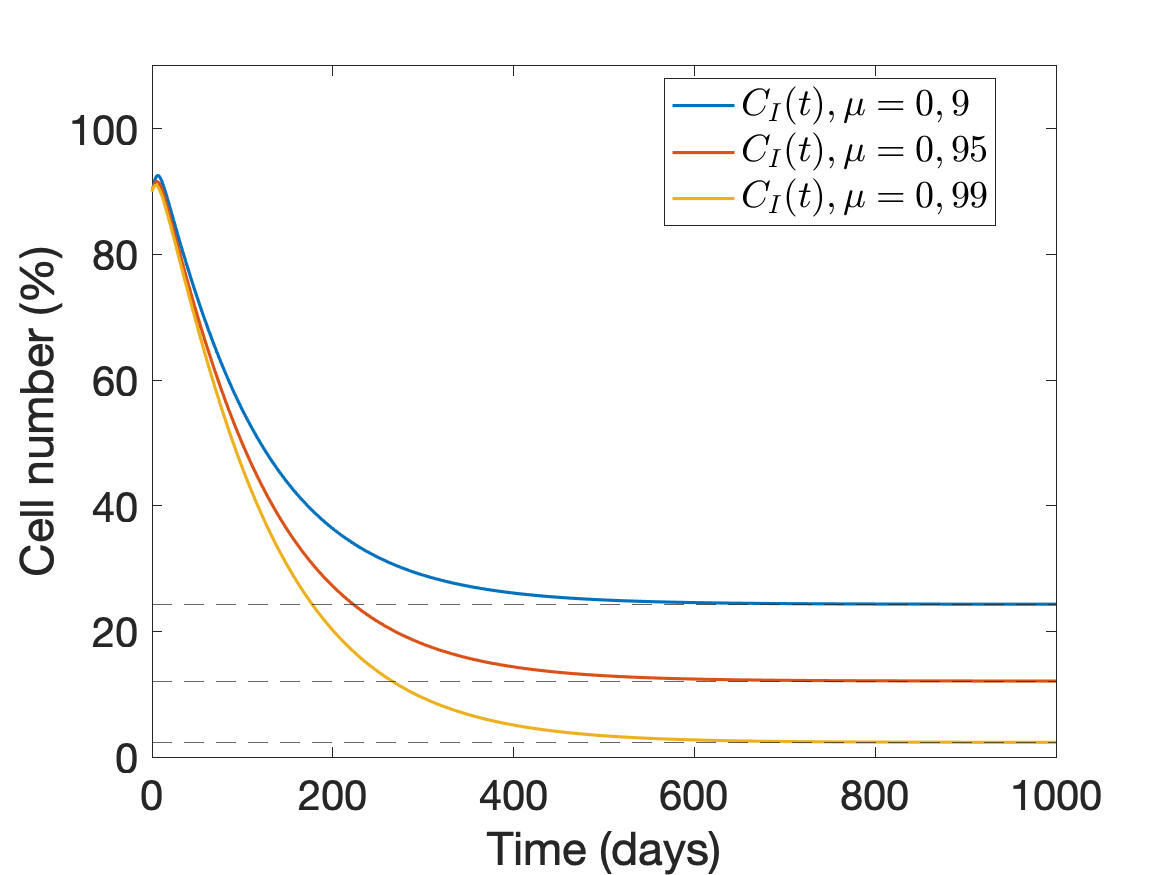} \\
 
    \caption{Analysis of the size of populations $P_m, P_p,$ and $C_I$ in terms of $\mu$, fixing the other parameters: $\rho = 0.15$ day$^{-1}$, $\rho_m = 0.3$ day$^{-1}$, $\alpha = 0.2$ day$^{-1}$, $\tau_s = 0.2$ day$^{-1}$,$\tau_c = 1/350000$ day$^{-1}$, $\gamma_c = 1/35$ day$^{-1}$, $\delta=100 $ day$^{-1}$, and $\delta_m=1$ day$^{-1}$.}
    \label{mu}
\end{figure}

Let us analyse the behavior of different populations in terms of $\mu$, for $\mu<1$.  When the populations associated to bone formation increase so that the load capacity, $M$, gets closer to $1$, both proliferation and differentiation decelerate. The closer to $1$ the parameter $\mu$ is, the more similar their decrease, and the faster differentiation decreases, of both WT and mutant SSCs. From our numerical simulations, we can also assert that the severity of the FD increases, because of the augmentation of mutant and WT phenocopying osteoprogenitors, and the diminution of osteocytes, see Figure \ref{mu}. This fact is consistent with the increased levels of RANKL in FD, which are thought to be responsible for the severity of the disease, \cite{deCastro2019}, and the fact that RANKL inhibits differentiation of SSCs into osteoprogenitors \cite{Chen2018}.
\newpage
\thispagestyle{empty}

\subsubsection{The unexpected low impact of the factual proliferation of mutant and WT phenocopying osteoprogenitors}

As it can be seen in Figure \ref{fig:Equilibrium_parameters} (e), the factual proliferation of mutant and WT phenocopying osteoprogenitors does not affect the equilibrium point. This fact is surprising due to the experimental evidence that cell proliferation is two-fold to threefold larger in osteoblastic cells expressing the mutation, compared with normal cells from the same patient, \cite{Marie1997}. Experiments also showed that it was also larger in cells isolated from more severe than less severe fibrotic lesions.

\section{Discussion}
\label{discussion}

Fibrous dysplasia is a rare bone disorder with an incidence estimated at 1 in 5.000 to 10.000 individuals, \cite{Pai2013}. The frequency of this condition in a general population is difficult to determine due to the fact that mild cases are usually not diagnosed. The small incidence and the diversity of clinical presentations makes it challenging to conduct large-scale clinical trials and develop effective treatments. In this context, theoretical modeling and in silico approaches can play an important role in understanding the disease mechanisms and developing optimal therapies. In the current century, several mathematical models of different aspects of bone biology have been developed to recapitulate and predict remodeling behavior, but to our knowledge none of them has addressed the study of fibrous dysplasia.
 
The development of quantitative approaches for studying fibrous dysplasia is important for advancing our understanding of the disease and for finding the best therapeutic combination regimens for specific patients. Overall, the development of a mathematical model of fibrous dysplasia is an important first step towards a more quantitative and predictive understanding of the disease. 

In this paper, we have constructed a mathematical model of remodeling dynamics in bone affected by fibrous dysplasia. Our model consists of a system of ordinary differential equations that describes the interactions between healthy bone-forming cells, mutant osteoprogenitor cells, WT phenocopying cells, mature bone cells, and bone-resorbing cells. To keep the number of independent elements and parameters to be fitted limited, the signaling that regulates the interaction between the cellular populations has been incorporated implicitly through direct interaction terms. 

Interestingly, taking biologically relevant parameter values, the system has a unique positive equilibrium point which is uniformly asymptotically stable and leads to an abnormal immature bone as the one resulting from the effect of the disease. When neither mutant progenitors nor mutant phenotype cells are incorporated, the model recapitulates the standard composition of the normal bone. 
Using this model, researchers could explore how changes in the parameters, for instance due to pharmacological actions, could affect the development and progression of fibrous dysplasia. 

Also, the model displays other interesting features providing a mechanistic support to
different clinical observations. One example of this is the fundamental role of the parameter measuring the flow of osteoprogenitors differentiated from mutant SSCs, which supports age-dependent normalization of FD lesions, due to the fact that mutant SSCs enter senescence before WT SSCs do, and their progeny is consumed by apoptosis, \cite{Kuznetsov2008}. To properly study this phenomenon, a specific model to consider the evolution of the lesions along the years of adulthood is required. In that model, the reversibility of the phenotype of WT phenocopying osteoprogenitors should be included.

The saturation parameter in the differentiation of both mutant and WT SSCs, which modulates the flow, has also a crucial role: the faster the flow decreases when the load capacity $M$ is near $1$, the more severe the FD. This agrees with the increased levels of RANKL in FD, \cite{deCastro2019}, and the fact that RANKL inhibits osteoblast differentiation \cite{Chen2018}. It is known that RANKL is overproduced in FD tissue, causing the number and the activity of osteoblasts to increase. Our results suggest that the overproduction of RANKL also contributes to the aggressiveness of FD lesions due to a modification in the differentiation of both mutant and WT SSCs cells. Specifically, the decrease of the flow of WT SSCs together with the overproduction of mutant and WT phenocopying osteoprogenitors can be responsible for the formation of the fibrous tissue.

The formation and activation of osteoclasts due to signaling from osteoprogenitors and osteocytes is the most important parameter for osteoclasts, being consistent with the active osteoclastogenesis in FD lesions \cite{Riminucci2003}. As for osteocytes, it is key the flow from stem cells, the mechanism closing that flow, as well as the osteoclasts activation and formation. Therefore, according to our model, the increased competence for space in the bone affects the osteocyte population, and the increase in the number of osteoclasts outperforms the augmentation of its activity. 

 On the contrary, a surprising observation relating the factual proliferation of mutant and WT phenocopying osteoprogenitors, which turned out to have a limited impact in our model, does not match the well-known aberrantly high proliferation of those cell populations \cite{Marie1997}. Those populations are also known to be highly apoptotic \cite{Kuznetsov2008}, does apoptosis compensate proliferation in these cell populations? Or the obtained results are telling us that contact inhibition is suppressed in these populations? 

Theoretical modeling could be a valuable tool for gaining insights into the mechanisms underlying fibrous dysplasia. Computational models can simulate the intricate interactions between various signaling pathways involved in bone growth and remodeling, shedding light on how alterations in these pathways can lead to the development of fibrous dysplasia. 

However, it is crucial to remark that models are rough approximations of the reality.  On the one hand, we do have the inherent limitations of ordinary differential equations modeling, such as the lack of spatial details of the solutions. On the other hand, many principles and variables are not fully accounted for. Otherwise, no information could be obtained from them due to its high complexity. In our proposed model, we made certain assumptions that may not capture the full complexity of the disease. For instance, the anabolic role of osteclasts in osteoprogenitors differentiation \cite{Bolamperti2022} has not been taken into account. Besides, the parameter measuring the efficiency of mutant osteoprogenitors to convert normal osteoprogenitors into WT phenocopying ones, which demonstrated to be crucial for the population of WT phenocopying osteoprogenitors, appears in a term designed to model the contact between two populations whose members are all in touch, but this is radically not the case. Even more, the percentage of WT osteoprogenitors which are in contact with mutant ones is thought to increase drastically over time. Therefore, in a more realistic (and complicated) model, that parameter should not be constant. Finally, the reversibility of that phenotype has not been considered, and it should be included to study the effect of medication on FD lesions.

Moving forward, our research aims to expand on these models in order to get an answer to the questions that have arisen. As well as to incorporate the effects of denosumab, a medication used in clinical trials as a potential treatment to fibrous dysplasia, \cite{deCastro2023}. By integrating denosumab into the model, we can assess its potential efficacy. Moreover, we can explore different treatment protocols and dosing schedules in order to avoid/minimize the rebound in bone turnover after discontinuation of denosumab treatment which has been reported \cite{Boyce2012,Collins2020,Meier2021}, as well as its combination with other therapies \cite{Makras2021}. This will allow us to optimize therapeutic strategies and determine the most effective doses for managing fibrous dysplasia. Addressing these open questions in fibrous dysplasia research is crucial for advancing our understanding of the disease and improving treatment outcomes. Exploring alternative approaches to describe bone formation processes, such as the simulation of a two-dimensional cellular automaton, we may offer additional insights and complement the existing modeling frameworks.

We hope this study stimulates the research in the fascinating area of bone disorders, and the specific problem of fibrous dysplasia.

\backmatter

\bmhead{Acknowledgments}
M.S. has been supported by University of Castilla-La Mancha (European Social Fund Regional Operative Program 2014-2020).
This research has been partially supported by Ministerio de Ciencia e Innovación, Spain (doi:10.13039/501100011033, grant number PID2022-142341OB-I00), Junta de Comunidades de Castilla-La Mancha (grant number SBPLY/21/180501/000145), and University of Castilla-La Mancha grant 2022-GRIN-34405 (Applied Science Projects within the UCLM research programme). M.C. has been partially supported by Spanish MICINN project PID2021-126217NB-I00 and by the European Union - NextGenerationEU program. 

\subsection*{Conflicts of interest} We declare we have no competing interests.

\subsection*{Data Availability}  Data sharing not applicable to this article as no datasets were generated or analysed during
the current study.

\subsection*{Authors' contributions}

\noindent M.S: Conceptualization, Methodology, Formal Analysis, Investigation, Writing-Review \& Editing. J.C.B.V: Conceptualization, Methodology. Writing-Review \& Editing
L.F. de C: Supervision, Writing-Review \& Editing. J.B.-B: Formal analysis, Writing-Review \& Editing. V.M.P-G: Conceptualization, Methodology, Supervision, Writing-Review \& Editing, Funding Acquisition. 
M.C: Conceptualization, Methodology, Supervision, Formal Analysis, Writing-Original Draft.

%\bibliography{sn-bibliography}


\end{document}